\documentclass[aps,prd,10pt,twocolumn,floatfix,superscriptaddress,nofootinbib]{revtex4-2}
\usepackage{amsmath,amssymb,graphicx,xcolor}
\usepackage{orcidlink}
\graphicspath{{figures/}}
\newcommand{\Eth}{E_{\rm th}}
\newcommand{\PU}{\affiliation{\small Phenikaa Institute for Advanced Study, Phenikaa University, Nguyen Trac, Duong Noi, Hanoi 100000, Vietnam}}
\newcommand{\NCTS}{\affiliation{\small Physics Division, National Center for Theoretical Sciences, National Taiwan University, Taipei 106319, Taiwan}}
\newcommand{\NTU}{\affiliation{\small Department of Physics and Center for Theoretical Physics, National Taiwan University, Taipei 10617, Taiwan}}

\begin{document}
\raggedbottom
\title{Threshold effects of inelastic neutrino--dark-matter scattering in AGN spikes}

\author{Cheng-Wei Chiang\orcidlink{0000-0003-1716-0169}}%
\email{chengwei@phys.ntu.edu.tw}\NTU \NCTS

\author{Van Que Tran\orcidlink{0000-0003-4643-4050}}%
\email{vqtran@phys.ncts.ntu.edu.tw} \NCTS \PU

\date{\today}
\begin{abstract}

High-energy neutrinos from active galactic nuclei provide a powerful probe of dark matter (DM) spikes surrounding supermassive black holes. While elastic neutrino--DM scattering mediated by a light particle produces little net attenuation due to regeneration, inelastic scattering, $\nu\chi_1\to\nu\chi_2$, introduces a nonzero minimum energy loss, $E_{\rm th}$.
We show that this threshold prevents the near-cancellation between attenuation and regeneration, producing a distinctive spectral deficit above $E_{\rm th}$ accompanied by a pile-up below it. Solving the full neutrino cascade, including repeated scattering and secondary neutrinos from excited-state decays, we demonstrate that this feature persists for both narrow and broad energy-transfer kernels. Using public IceCube data for NGC~1068 and an unbinned likelihood analysis with a freely fitted intrinsic power-law spectrum, we derive upper limits on the asymptotic optical depth of $\tau_0<26$--39 for a narrow kernel (kernel width $\epsilon=0.01$) over $E_{\rm th}=0.24$--3.0~TeV and $\tau_0<20$--26 for a broad kernel ($\epsilon=0.36$) over $E_{\rm th}=0.24$--2.4~TeV, at the nominal 90\% confidence level. An intrinsic high-energy cutoff weakens the narrow-kernel limits by factors of 2.2--7.4 and the broad-kernel limits by factors of 2.0--5.0. We further investigate DM velocity effects and combine NGC~1068 with CGCG~420-015 and TXS~0506+056 to extend the threshold reach. In a vector-mediator realization, an optimistic adiabatic DM spike can yield constraints on the neutrino coupling competitive with laboratory bounds for broad energy-transfer kernels.
\end{abstract}
\maketitle

\section{Introduction}\label{sec:intro}

IceCube has reported evidence for neutrino emission 
from the Seyfert galaxy NGC~1068~\cite{IceCube:2022der,IceCube:2019cia,Abbasi:2026ehs} 
and the blazar TXS~0506+056~\cite{IceCube:2018cha}. 
If dark matter (DM) forms a dense spike around the central supermassive black hole~\cite{Gondolo:1999ef,Vasiliev:2007vh,Shapiro:2016ypb,Shapiro:2022prq}, 
neutrinos emitted from these sources may traverse substantial DM column densities, 
$\Sigma\equiv\int\rho_\chi\,d\ell\sim10^{29}$--$10^{32}$~GeV\,cm$^{-2}$, 
where $\rho_\chi$ is the DM density and the integral extends along the neutrino trajectory from the emission region. 
The observed neutrino fluxes can therefore place stringent constraints on neutrino--DM scattering~\cite{Arguelles:2017atb,Kelly:2018tyg,Choi:2019ixb,Murase:2019xqi,Cline:2022qld,Cline:2023tkp,Ferrer:2022kei,Dixit:2026zkv,Fujiwara:2024qos,Zapata:2025huq,Tseng:2024akh,Esteban:2025wbv}. 
However, scattering primarily redistributes neutrinos in energy rather than removing them from the flux, 
making the resulting constraints sensitive to the energy transferred in each scattering event. 

For \emph{elastic} scattering mediated by a light $t$-channel particle, 
the energy transfer is typically small. Scattered neutrinos therefore retain nearly their original energies, 
and the flux depleted at a given energy is largely replenished 
by neutrinos scattered down from slightly higher energies. 
This regeneration effect strongly suppresses the net attenuation~\cite{Farzan:2026wek} (see also~\cite{Cline:2026stf}).
In contrast, \emph{inelastic} scattering that excites the DM ground state 
$\chi_1$ to a heavier state $\chi_2$, $\nu\chi_1\to\nu\chi_2$, 
involves a minimum neutrino energy loss of $\Eth=(m_2^2-m_1^2)/(2m_1)$, 
where $m_1$ and $m_2$ are the masses of $\chi_1$ and $\chi_2$, respectively. 
This nonzero minimum persists even in the forward-scattering limit, 
preventing the near-cancellation between attenuation 
and regeneration characteristic of elastic scattering. 
Consequently, the transmission, defined as the ratio of 
the neutrino flux emerging from the spike to the intrinsic source flux at the same energy, 
develops a distinctive spectral feature: 
a deep deficit just above $\Eth$ accompanied by a pile-up just below it, 
as illustrated in Fig.~\ref{fig:f}(c).

The threshold itself is standard kinematics~\cite{Tucker-Smith:2001myb}. 
What distinguishes this work from previous studies of neutrino--DM attenuation 
in active galactic nuclei (AGN) spikes is the nonzero minimum energy transfer: 
those studies assume absorption, an $\mathcal O(1)$ fractional energy loss (heavy mediators or isotropic
scattering~\cite{Cline:2022qld,Tseng:2024akh}), or an elastic kernel extending to zero energy transfer~\cite{Farzan:2026wek}.
We (i) apply the inelastic threshold to neutrino propagation through AGN spikes; 
(ii) solve the full energy-space cascade with repeated scattering,
secondary neutrinos from $\chi_2$ decay, and the DM velocity distribution; 
and (iii) derive the observable spectral distortion and the resulting IceCube constraints.
 Inelasticity on the neutrino side was studied in Ref.~\cite{Abbaslu:2026row}; 
 there the heavy state carries almost the full
neutrino energy, so the spectrum keeps its shape and the signal is in the flavor ratios. 
Inelastic DM in AGN has so far been probed with cosmic
rays~\cite{Gustafson:2024aom,Gustafson:2025dff,Chun:2026muh}.

Our results can be interpreted at three levels, 
with progressively increasing dependence on astrophysical assumptions. 
First, the nonzero minimum energy loss is determined by stationary-target kinematics, 
while the resulting spectral edge depends on the energy-transfer kernel and the intrinsic source spectrum. 
Second, constraining the asymptotic optical depth $\tau_0$, 
defined as the optical depth in the limit $E_\nu\gg\Eth$ 
and encoding the product of the scattering cross section and DM column density, 
additionally requires modeling the source spectrum and detector response. 
Finally, translating the optical-depth constraint into bounds on 
the underlying particle-physics parameters requires assumptions 
about the DM spike profile and a specific particle-physics model.

The paper is organized as follows.
Section~\ref{sec:threshold} derives the kinematic threshold.
Section~\ref{sec:kernel} introduces the phenomenological energy-transfer kernel and the transport equation,
and Sec.~\ref{sec:cascade} presents the full cascade computation, including angular deflection and DM velocity effects.
Section~\ref{sec:spike} describes the DM spike benchmarks.
In Sec.~\ref{sec:ngc}, we derive limits on $\tau_0$ from IceCube observations of NGC~1068,
and in Sec.~\ref{sec:other}, we extend the analysis to CGCG~420-015 and TXS~0506+056.
Section~\ref{sec:model} interprets the limits in a vector-mediator model,
and Sec.~\ref{sec:conclusions} summarizes our conclusions.
The particle-physics model, the numerical transport, the multi-source analysis, the bounds on $g_\chi$,
and the dependence of the limits on the kernel width are detailed in Appendices~\ref{app:model}--\ref{app:kernelwidth}.

\section{Kinematic threshold}
\label{sec:threshold}

Consider the inelastic scattering process $\nu(k)+\chi_1(p)\to\nu(k')+\chi_2(p')$, 
with the initial DM particle $\chi_1$ at rest. 
Imposing the on-shell condition $p'^2=m_2^2$ yields the exact relation
\begin{equation}
\omega\equiv E_\nu-E_\nu'
=\frac{m_2^2-m_1^2+Q^2}{2m_1}
\ge \Eth\equiv\frac{m_2^2-m_1^2}{2m_1},
\label{eq:omega}
\end{equation}
where $E_\nu$ ($E_\nu'$) is the incoming (outgoing) neutrino energy, 
and $Q^2=-t\ge0$, with $t=(k-k')^2$ denoting the Mandelstam variable. 
Throughout this work, we adopt natural units, $\hbar=c=1$, unless stated otherwise.

Equation~\eqref{eq:omega} establishes a nonzero minimum energy loss, $\omega_{\min}=\Eth$, 
attained in the forward-scattering limit $Q^2=0$. 
Unlike in elastic scattering, a vanishing momentum-transfer invariant $Q^2$ 
does not imply vanishing energy transfer, since part of 
the incident neutrino energy is required to excite the DM particle into the heavier state $\chi_2$. 
It is therefore important to distinguish the kinematic minimum $\omega_{\min}$, 
the energy loss in an individual scattering event, $\omega$, 
and the mean energy loss $\langle\omega\rangle$. 
While the minimum energy loss is fixed entirely by kinematics, 
the width and shape of the energy-loss distribution depend on the underlying interaction.

\section{Kernel and transport}\label{sec:kernel}

We adopt a phenomenological parametrization of the inelastic scattering cross section, 
characterized by its normalization, kinematic threshold, and kernel width. 
A concrete particle-physics realization is presented in Sec.~\ref{sec:model} and Appendix~\ref{app:model}. 
In particular, we consider the following parametrization of the total and differential cross sections:
\begin{equation}
\label{eq:sigma}
\begin{aligned}
\sigma(E_\nu)&=\sigma_0\left(1-\frac{\Eth}{E_\nu}\right)\Theta(E_\nu-\Eth),\\
\frac{d\sigma(E_\nu\to E_\nu')}{dE_\nu'}&=
\frac{\sigma(E_\nu)}{\mathcal N}
\frac{\epsilon\Eth\;\Theta(E_\nu-E_\nu'-\Eth)}
{\left(E_\nu-E_\nu'-\Eth+\epsilon\Eth\right)^2},
\end{aligned}
\end{equation}
for $E_\nu'\ge0$, where $\Theta$ is the Heaviside step function. 
Here, $\sigma_0$ denotes the asymptotic high-energy cross section, 
while the dimensionless kernel width $\epsilon$ controls 
the characteristic width of the energy-loss distribution above the kinematic minimum $\Eth$. 
The normalization factor $\mathcal N=(E_\nu-\Eth)/(E_\nu-\Eth+\epsilon\Eth)$ 
ensures that $\int dE_\nu'\,d\sigma/dE_\nu'=\sigma(E_\nu)$. 
We refer to the normalized distribution of outgoing neutrino energies 
as the energy-transfer kernel, discussed in the following section, and $\epsilon$ is defined as its width.
For a narrow kernel width ($\epsilon\ll1$), 
energy losses are concentrated near $\Eth$, 
with a characteristic spread of order $\epsilon\Eth$ 
and a power-law tail proportional to $(\omega-\Eth)^{-2}$ at larger energy transfers.
We do not restrict our analysis to the narrow-kernel regime. 
Instead, we consider two benchmark values of the kernel width: 
$\epsilon=0.01$, representing a narrow kernel with energy losses concentrated near the kinematic threshold, 
and $\epsilon=0.36$, corresponding to a broader energy-loss distribution.

Equation~\eqref{eq:sigma} approximates the high-energy behavior of 
inelastic scattering mediated by a $t$-channel vector boson of mass $m_{Z'}$, 
for which $\epsilon=m_{Z'}^2/(m_2^2-m_1^2)$. 
In our numerical analysis, we employ the full cross section 
and energy-transfer kernel derived from this vector-mediated interaction, 
evaluated at the same $\Eth$ and $\epsilon$. 
Equation~\eqref{eq:sigma} serves to define the phenomenological parameters. 
We have verified that replacing the full expressions 
with Eq.~\eqref{eq:sigma} changes the event-count limits 
by at most 6\% for $\epsilon=0.01$, but by 11--50\% for $\epsilon=0.36$ at $\Eth=0.1$--2.5~TeV.
For the broad kernel, the full cross section reaches the asymptotic form of Eq.~\eqref{eq:sigma} only for $(E_\nu-\Eth)/\Eth\gg\epsilon$.
Equation~\eqref{eq:sigma} therefore overestimates the attenuation just above threshold.
Apart from this comparison, the sharp-shift approximation in this section, and the comparison in Appendix~\ref{app:model}, 
all results presented below, 
for both narrow and broad energy-transfer kernels, are obtained using the full expressions.

The energy-dependent optical depth is given by
\begin{equation}
\tau(E_\nu)=\sigma(E_\nu)X_{\rm tot}
\simeq\tau_0\left(1-\frac{\Eth}{E_\nu}\right),
\label{eq:tau}
\end{equation}
where $X_{\rm tot}\equiv\Sigma/m_1$ is the total $\chi_1$ number column density, 
in units of cm$^{-2}$, 
and $\tau_0\equiv\sigma_0X_{\rm tot}$ denotes the dimensionless asymptotic optical depth. 
The approximation in Eq.~\eqref{eq:tau} becomes exact for the parametrization in Eq.~\eqref{eq:sigma}. 
We emphasize that $\tau_0$ characterizes the interaction probability evaluated using the asymptotic cross section 
and should not be identified with the actual number of scatterings experienced by a neutrino, 
whose energy and hence scattering probability evolve with propagation.

The phenomenological parameter space is therefore specified by $\tau_0$, $m_1$, $\Eth$, and $\epsilon$, 
with $\tau_0$ encoding the product of the scattering cross section and DM column density. 
We present our constraints as upper limits on $\tau_0$ as functions of $\Eth$ and $\epsilon$. These limits are converted into bounds on $\sigma_0$ in Sec.~\ref{sec:other} and on the couplings of the vector-mediator model in Sec.~\ref{sec:model}.

We define the traversed $\chi_1$ number column density as 
$X\equiv\int n_1\,d\ell$, where $n_1=\rho_\chi/m_1$ is the DM number density 
and $0\le X\le X_{\rm tot}$, with $X$ measured in cm$^{-2}$. 
Let $\Phi(E_\nu;X)$ denote the differential neutrino flux (hereafter simply the flux), 
summed over all three neutrino and antineutrino flavors, 
at column depth $X$, where $E_\nu$ is the neutrino energy in the source frame. 
The energy-independent geometric dilution along the line of sight is divided out, 
so $\Phi$ changes only through scattering on $\chi_1$ 
and through the secondary neutrinos from the prompt decays of the $\chi_2$ it produces. 
Neutrino oscillations redistribute the flux among flavors but leave the flavor-summed flux unchanged.

We define the intrinsic (emitted) and transmitted fluxes as $\Phi^{\rm in}(E_\nu)\equiv\Phi(E_\nu;0)$ and $\Phi^{\rm out}(E_\nu)\equiv\Phi(E_\nu;X_{\rm tot})$, respectively. The energy-dependent transmission factor is then defined as
\begin{equation}
R(E_\nu;\tau_0,\Eth,\epsilon)\equiv
\frac{\Phi^{\rm out}(E_\nu)}
{\Phi^{\rm in}(E_\nu)}.
\label{eq:R}
\end{equation}
Because the transport equation is linear in the neutrino flux, $R$ is independent of the overall normalization of the intrinsic spectrum. However, regeneration couples different neutrino energies, making $R$ sensitive to the spectral shape. For the intrinsic power-law spectra $\Phi^{\rm in}\propto E_\nu^{-\gamma}$ considered throughout this work, $R$ therefore also depends on the spectral index $\gamma$, which is left implicit in our notation.

In the particle-physics model of Sec.~\ref{sec:model}, the neutrino coupling $g_\nu$ is flavor universal, yielding identical scattering cross sections for all neutrino and antineutrino flavors. The secondary neutrinos from $\chi_2$ decays are likewise produced with equal flavor branching fractions. The flavor-summed flux therefore obeys a closed transport equation, and $R$ is independent of the flavor composition of the source. The composition enters only through the $\nu_\mu+\bar\nu_\mu$ fraction at Earth, up to a correction of less than 10\% from the flavor-democratic secondaries, provided that the source composition does not vary with energy. In NGC~1068, the neutrinos have been proposed to originate from photohadronic ($p\gamma$) and hadronuclear ($pp$) interactions of protons accelerated in the X-ray corona~\cite{Murase:2019vdl,Inoue:2019yfs}. The resulting pion decays give the source flavor ratio $\nu_e:\nu_\mu:\nu_\tau=1:2:0$, for which averaged oscillations give a muon-flavor fraction of 0.32--0.37 at Earth over the $3\sigma$ ranges of the mixing angles~\cite{Esteban:2024eli} and any CP phase, close to the fraction 1/3 of the flavor-democratic secondaries. We therefore adopt a flavor-democratic flux with $\nu:\bar\nu=1:1$. A flavor-dependent coupling would instead make $R$ flavor dependent and could induce additional distortions of the $\nu_\mu+\bar\nu_\mu$ spectrum.

The evolution of the neutrino flux $\Phi(E_\nu;X)$ through the DM spike is governed by the transport equation
\begin{equation}
\label{eq:dflux}
\begin{aligned}
\frac{d\Phi(E_\nu;X)}{dX}={}&
-\sigma(E_\nu)\Phi(E_\nu;X)\\
&\hspace{-1.5cm} +\int_{E_\nu}^{\infty}d\widetilde E_\nu\,
\sigma(\widetilde E_\nu)
\left[K+K_{\rm sec}\right]
(\widetilde E_\nu\to E_\nu)\,
\Phi(\widetilde E_\nu;X),
\end{aligned}
\end{equation}
where $\widetilde E_\nu > E_\nu$ and 
\begin{equation}
\label{eq:kernels}
\begin{aligned}
K(\widetilde E_\nu\to E_\nu) &= \frac{1}{\sigma(\widetilde E_\nu)} \frac{d\sigma(\widetilde E_\nu\to E_\nu)}{dE_\nu}, \\
K_{\rm sec}(\widetilde E_\nu\to E_\nu)&=
\int_{\Eth}^{\widetilde E_\nu}d\omega\,
K(\widetilde E_\nu\to\widetilde E_\nu-\omega)\\
&\;\;\times\frac{1}{m_1+\omega}\,
f_{\nu\bar\nu}\left(\frac{E_\nu}{m_1+\omega}\right).
\end{aligned}
\end{equation}
Here $K$ is the normalized energy-transfer kernel, 
describing the distribution of outgoing neutrino energies after a single scattering. 
The secondary-neutrino kernel $K_{\rm sec}$ accounts for secondary neutrinos produced 
in the subsequent decay of the excited DM state $\chi_2$. 
For a scattering event with energy loss $\omega$, 
the outgoing $\chi_2$ carries an energy $E_{2}=m_1+\omega$ 
and is assumed to decay promptly into $\chi_1$ and a $\nu\bar\nu$ pair. 
The secondary-neutrino energy distribution is described by $f_{\nu\bar\nu}(E_\nu/E_{2})$, 
accounting for both the neutrino and antineutrino produced in each decay (see Appendix~\ref{app:model}).

The first term on the right-hand side of Eq.~\eqref{eq:dflux} describes the depletion of neutrinos 
at energy $E_\nu$ due to scattering, 
while the second term accounts for the repopulation of the spectrum through 
downscattering of higher-energy neutrinos $\widetilde E_\nu$. 
This gain term includes both regeneration of the scattered neutrinos, 
described by $K$, and the additional neutrinos from $\chi_2$ decays, 
described by $K_{\rm sec}$. The kernels satisfy $\int dE_\nu\,K=1$ and $\int dE_\nu\,K_{\rm sec}=2$, 
reflecting the fact that each scattering preserves the incident neutrino 
and produces two additional neutrinos through the subsequent decay. 
Both contributions are included in all results, except for the primary-only comparison with the sharp-shift approximation below.
We emphasize that the secondary-neutrino contribution depends on the decay spectrum specified 
in Appendix~\ref{app:model} and is therefore not determined by the inelastic scattering threshold alone.

\begin{figure}[tbp]
\centering
\includegraphics[width=\columnwidth]{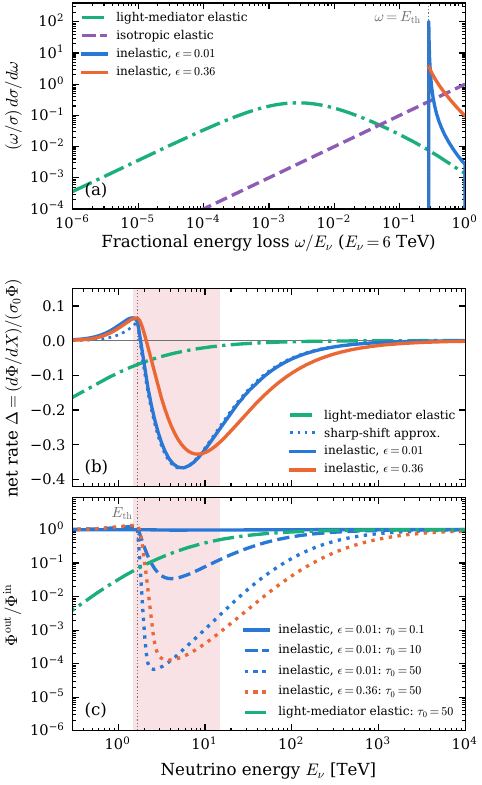}
\caption{
(a) Distribution of the fractional energy loss $\omega/E_\nu$ per scattering at $E_\nu=6$~TeV for light-mediator elastic scattering  (green dash-dotted), isotropic elastic scattering (purple dashed), and inelastic scattering with $\epsilon=0.01$ (blue) and 0.36 (orange).
(b) Net fractional flux change $\Delta$ for an intrinsic power-law spectrum $\Phi^{\rm in}\propto E_\nu^{-3.2}$, comparing light-mediator elastic scattering (green dash-dotted) with inelastic scattering including secondary neutrinos for $\epsilon=0.01$ (blue solid) and $\epsilon=0.36$ (orange solid), and with the sharp-shift approximation of Eq.~\eqref{eq:Delta_sharp} (blue dotted).
(c) Transmission factor $R=\Phi^{\rm out}/\Phi^{\rm in}$ obtained from the full cascade for inelastic scattering at $\tau_0=0.1$ (blue solid), 10 (blue dashed), and 50 (blue dotted) for $\epsilon=0.01$, and at $\tau_0=50$ for $\epsilon=0.36$ (orange dotted), compared with light-mediator elastic scattering at $\tau_0=50$ (green dash-dotted).
For the inelastic cases, we adopt $\Eth=1.67$~TeV ($m_1=300$~keV, $m_2=1$~GeV), 
$m_{Z'}=0.1$~GeV ($\epsilon=0.01$) or 0.6~GeV ($\epsilon=0.36$), and stationary $\chi_1$ targets. 
For light-mediator elastic scattering, we take the DM mass $m_{\rm DM}=300$~keV and the mediator mass $m_{\rm med}=0.1$~GeV.
Shaded: 1.5--15~TeV, the range contributing 68\% of the IceCube NGC~1068 test statistic~\cite{IceCube:2022der}.
}
\label{fig:f}
\end{figure}

The contrasting energy-loss distributions are illustrated 
in Fig.~\ref{fig:f}(a). At $E_\nu=6$~TeV, the mean fractional energy loss is 
$\langle\omega\rangle/E_\nu=1.2\times10^{-2}$ for light-mediator 
elastic scattering%
\footnote{For elastic scattering, we consider a Dirac fermion DM $\psi$ interacting with a vector mediator $Z'$ through the coupling $g_\chi Z'_\mu\bar\psi\gamma^\mu\psi$, which is the $m_2\to m_1$ limit of Eq.~\eqref{a:eq:L}.} with mediator mass $m_{\rm med}=0.1$~GeV and DM mass $m_{\rm DM}=300$~keV (it is $1.2\times10^{-3}$ for the benchmark $m_{\rm med}=\sqrt{2}$~GeV and $m_{\rm DM}=1$~GeV used in Ref.~\cite{Farzan:2026wek}), 
compared with 0.5 for isotropic elastic scattering mediated by a heavy particle,
for which we take $d\sigma/d\omega$ constant for $0\le\omega\le E_\nu$.
For inelastic scattering, we find $\langle\omega\rangle/E_\nu=0.29$ for $\epsilon=0.01$ (0.39 for $\epsilon=0.36$), 
close to the kinematic minimum $\Eth/E_\nu\simeq0.28$. 
As shown below, this difference allows a relatively narrow 
inelastic energy-loss distribution to produce substantial spectral attenuation, 
whereas strong regeneration suppresses the corresponding elastic light-mediator signal.

To quantify the local impact of scattering on the neutrino spectrum, 
we define the net fractional flux change per unit column depth, 
normalized by the asymptotic cross section $\sigma_0$, as
\begin{equation}
\Delta(E_\nu)\equiv
\frac{1}{\sigma_0\Phi(E_\nu)}
\frac{d\Phi(E_\nu)}{dX}.
\label{eq:Delta}
\end{equation}
Here, $\Delta<0$ indicates net flux depletion, 
while $\Delta>0$ corresponds to net enhancement. 
Figure~\ref{fig:f}(b) shows $\Delta$ at the entrance of the DM column ($X=0$), 
assuming an intrinsic power-law spectrum $\Phi^{\rm in}\propto E_\nu^{-3.2}$ 
and benchmark parameters $m_1=300$~keV, $m_2=1$~GeV (i.e., $\Eth=1.67$~TeV), and $\epsilon=0.01$, together with the broad-kernel benchmark $\epsilon=0.36$. 
These benchmark values are used for all spectra shown in Secs.~\ref{sec:kernel}--\ref{sec:ngc}. The limits are scanned over $\Eth$ (equivalently $m_2$), and over $\epsilon$ in Appendix~\ref{app:kernelwidth}, with $m_1=300$~keV kept fixed throughout. 

For stationary DM targets, the cross section normalization 
and column density enter the transport only through the asymptotic optical depth $\tau_0$. 
In the relativistic regime considered here, the dependence on $m_1$ at fixed $\Eth$ and $\epsilon$ is weak, 
since $m_1+\omega\simeq\omega$ in the secondary-neutrino spectrum.

For elastic scattering mediated by a light particle, depletion and regeneration largely cancel.  
In particular, for $m_{\rm DM}=300$~keV and $m_{\rm med}=0.1$~GeV, we obtain $\Delta=-0.09$, $-0.05$, and $-0.02$ at $E_\nu=1$, 3, and 10~TeV, 
varying smoothly with energy and without spectral structure (green dash-dotted curve in Fig.~\ref{fig:f}(b)). 
In contrast, inelastic scattering produces substantial net depletion 
because the nonzero minimum energy loss prevents this cancellation, as illustrated by the blue curve in Fig.~\ref{fig:f}(b). 
Specifically, we obtain $\Delta=-0.20$, $-0.37$, $-0.24$, and $-0.03$ at $E_\nu/\Eth=1.5$, 3, 10, and 100, respectively, for $\epsilon=0.01$ (blue curve). For $\epsilon=0.36$, the depletion is weaker just above threshold and extends to higher energies, $\Delta=-0.07$, $-0.29$, $-0.28$, and $-0.06$ at the same energies, with the strongest depletion, $\Delta=-0.33$, at $E_\nu\simeq5\Eth$ (orange curve). Because each scattering removes more energy on average, regeneration from higher energies compensates less of the loss far above threshold. 
Below the inelastic threshold, the depletion term vanishes, 
while regeneration from higher energies enhances the neutrino flux, giving rise to a characteristic spectral pile-up.

The excited DM state subsequently undergoes prompt decay, $\chi_2\to\chi_1\nu\bar\nu$, 
mediated by an on-shell particle in our model.
The secondary neutrinos carry approximately $50\%$ ($\epsilon=0.01$) to 68\% ($\epsilon=0.36$)
of the excited-state energy in each scattering event. 
Since their total energy is bounded by the excited-state energy $m_1+\omega$, 
which is typically of order $\Eth$ for both kernels considered here ($\langle\omega\rangle\le1.42\,\Eth$ at $E_\nu=6$~TeV), 
the secondary contribution concentrates predominantly below the inelastic threshold. 
Consequently, secondary neutrinos enhance the spectral pile-up, 
increasing $\Delta(0.9\Eth)$ from 0.041 to 0.065 for $\epsilon=0.01$; for $\epsilon=0.36$, $\Delta(0.9\Eth)=0.063$. Above threshold, their effect is negligible for the narrow kernel. For the broad kernel, it is comparable to the net rate just above threshold, where it changes $\Delta$ by about $2\times10^{-2}$ and sets its sign, and it falls below a few percent of $\Delta$ above ${\sim}3\Eth$.

The underlying mechanism can be understood analytically 
using a simplified sharp-shift approximation, 
in which each scattering transfers exactly the minimum energy $\Eth$, 
and secondary neutrinos are neglected. 
The energy-transfer kernel then reduces 
to $K(\widetilde E_\nu\to E_\nu)=\delta(E_\nu-\widetilde E_\nu+\Eth)$.  
By construction, this approximation applies to narrow kernels, $\epsilon\ll1$, for which the energy loss is concentrated near $\Eth$. 
For an intrinsic power-law spectrum $\Phi^{\rm in}\propto E_\nu^{-\gamma}$ and the asymptotic cross section $\sigma=\sigma_0(1-\Eth/E_\nu)$, 
Eq.~\eqref{eq:Delta} evaluated at the entrance of the DM column becomes 
\begin{equation}
\Delta_{\rm ss}(E_\nu)={}
-\left(1-\frac{\Eth}{E_\nu}\right)
\Theta(E_\nu-\Eth)
+ \left(\frac{E_\nu}{E_\nu+\Eth}\right)^{1 + \gamma}.
\label{eq:Delta_sharp}
\end{equation}
The first term describes the depletion of neutrinos at energy $E_\nu$, 
while the second accounts for regeneration from neutrinos initially at $E_\nu+\Eth$. 

The approximation reproduces the corresponding primary-only 
numerical results for $\Delta$ to within 8\% over $2\le E_\nu/\Eth\le10^3$.  Since for $\epsilon=0.01$ secondaries change $\Delta$ by only ${\sim}10^{-3}$ above $1.2\Eth$, the approximation can be compared there directly with the full result, as shown by the blue dotted and solid curves in Fig.~\ref{fig:f}(b). 
Closer to threshold, deviations become more pronounced because 
the finite width of the energy-transfer distribution 
is increasingly important. 
In particular, the deviations are approximately $6\%$ and $18\%$ 
at $E_\nu/\Eth=1.5$ and 1.2, respectively.
Below threshold, the sharp-shift approximation 
reproduces the primary-only regeneration contribution to within 8\% for $E_\nu\ge0.18\Eth$.
For $\epsilon=0.36$, on the other hand, the mean loss exceeds $\Eth$ by about 40\%, and the sharp-shift approximation describes the mechanism only qualitatively.  

\section{Full cascade}
\label{sec:cascade}

The transmission factor $R(E_\nu)$ describes the cumulative modification 
of the neutrino spectrum after repeated scattering through the DM spike. 
The corresponding observable event distribution is obtained 
by convolving the transmitted spectrum with the detector response. 
To determine $R(E_\nu)$, we solve Eq.~\eqref{eq:dflux} over the full DM column 
using a logarithmic energy grid, with analytically 
integrated bin-to-bin transition kernels and matrix exponentiation. 
We validate our numerical implementation against an independent 
event-by-event Monte Carlo simulation employing the same scattering cross section, 
finding agreement within the Monte Carlo statistical uncertainty of a few percent. 
The solver also converges toward the results of $\nu$FATE~\cite{Vincent:2017svp} 
as the energy-node spacing is refined (see Appendix~\ref{app:transport}). 
As an additional cross-check, our implementation reproduces 
the NGC~1068 and TXS~0506+056 analyses of Refs.~\cite{Cline:2023tkp,Cline:2022qld} 
from their stated inputs.

The resulting transmission spectra are shown in Fig.~\ref{fig:f}(c). 
For an asymptotic optical depth of $\tau_0=50$, elastic scattering mediated 
by a light particle yields $R=0.14$ at 3~TeV,
whereas inelastic scattering with a small kernel width suppresses the transmission to $R=8\times10^{-5}$ at the same energy. 
This pronounced difference reflects the incomplete cancellation between attenuation
and regeneration caused by the nonzero minimum energy loss.
The inelastic transmission exhibits a characteristic minimum 
at $E_\nu\simeq1.6\Eth$ for $\tau_0=50$, shifting to $E_\nu\simeq2.5\Eth$ 
for $\tau_0=10$. 
Increasing the kernel width $\epsilon$ 
shifts the minimum toward higher energies, 
because the energy-transfer distribution broadens as shown in Fig.~\ref{a:fig:xs}(b) of Appendix~\ref{app:model}. 
For $\epsilon=0.36$ and $\tau_0=50$, the minimum moves from $1.6\Eth$ to $2.3\Eth$, 
with $R=1.3\times10^{-4}$ instead of $6.8\times10^{-5}$. 
The transmission just above threshold is larger, $R(1.5\Eth)\simeq1\times10^{-3}$ instead of $6.8\times10^{-5}$, 
whereas the recovery at higher energies is slower, $R(10\Eth)=1.1\times10^{-3}$ instead of $3.8\times10^{-3}$, 
as shown by the orange curve in Fig.~\ref{fig:f}(c).

The scattering-induced angular deflection is estimated as 
$\theta\simeq\sqrt{Q^2/(E_\nu E_\nu')}$, 
with root-mean-square values of $0.010^\circ$ ($0.047^\circ$) per scattering at $E_\nu=3$~TeV
for $\epsilon=0.01$ (0.36). 
A kernel-independent bound follows from Eq.~\eqref{eq:omega}: 
since $Q^2=2m_1(\omega-\Eth)\le2m_1(E_\nu-E_\nu')$, 
each scattering gives $\theta^2\le2m_1(1/E_\nu'-1/E_\nu)$. 
Summed over successive scatterings, this bound telescopes, 
so a neutrino observed at energy $E_\nu$ has accumulated a root-mean-square deflection 
below $(2m_1/E_\nu)^{1/2}$, i.e., $0.036^\circ$ at $E_\nu=1.5$~TeV for $m_1=300$~keV, 
for any kernel width and number of scatterings. 
This is well below the typical IceCube track 
angular resolution of $\gtrsim0.2^\circ$. Angular redistribution can 
therefore be neglected, allowing us to describe 
neutrino propagation using an energy-space transport equation alone.
This holds for both kernels:
a Monte Carlo simulation of successive scatterings gives root-mean-square deflections of at most 29\% ($\epsilon=0.01$)
and 74\% ($\epsilon=0.36$) of the bound, where the maximum is taken over observed energies and optical depths.

Our baseline calculation assumes stationary DM targets. 
Under this approximation, the transport depends on the DM distribution 
only through the integrated column density, 
or equivalently the asymptotic optical depth $\tau_0$, 
and is independent of the detailed spatial profile along the neutrino trajectory. 
In a realistic DM spike, however, the particles have nonzero velocities, 
potentially reaching the local escape speed of $v_{\rm esc}\simeq(R_S/r)^{1/2}\simeq0.18c$ 
at $r=30R_S$, where $R_S=2GM_{\rm BH}$ is the Schwarzschild radius of the central black hole 
with mass $M_{\rm BH}$. 
For a moving target, the incident neutrino energy is Doppler shifted 
in the DM rest frame, effectively replacing the stationary-target threshold 
 $\Eth$ by $\Eth/D$, where $D$ is the corresponding Doppler factor:
a neutrino of lab energy $E_\nu$ has energy $DE_\nu$ in the rest frame of a $\chi_1$ with speed $v$ and Lorentz factor $\gamma_\chi$,
where $D=\gamma_\chi(1-v\cos\theta_\chi)$ and $\theta_\chi$ is the angle between the $\chi_1$ velocity and the neutrino direction,
so the threshold condition $DE_\nu\ge\Eth$ becomes $E_\nu\ge\Eth/D$.
Incorporating the DM velocity distributions associated with 
the spike models of Sec.~\ref{sec:spike}, 
we find, for $\epsilon=0.01$, that the spectral edge shifts by approximately 2--6\% 
and becomes broader, but remains clearly identifiable.
The resulting likelihood limits change by at most $2\%$,
indicating that the predicted spectral feature
and derived constraints are relatively insensitive
to DM motion.
The same holds for $\epsilon=0.36$: the edge shifts by 2--4\% and broadens,
and the event-count limits change by at most 7\% for $\Eth\le2.5$~TeV (5\% for $\epsilon=0.01$).

\section{Dark matter around supermassive black holes}
\label{sec:spike}

The adiabatic growth of a supermassive black hole 
at the center of a DM halo with an inner density cusp $\rho_\chi\propto r^{-\gamma_c}$ 
can induce a steep DM spike, $\rho_\chi\propto r^{-\alpha}$, with $\alpha=(9-2\gamma_c)/(4-\gamma_c)$. 
For a Navarro--Frenk--White (NFW) profile ($\gamma_c=1$), 
this yields $\alpha=7/3$~\cite{Gondolo:1999ef,Navarro:1996gj}. 
The spike extends from an inner cutoff at $4R_S$ 
(or $2R_S$ in a relativistic treatment~\cite{Sadeghian:2013laa}) 
to a characteristic radius $R_{\rm sp}$, beyond which it matches the host halo profile. 
However, the spike density is subject to substantial astrophysical uncertainties. 
DM annihilation can flatten the inner density profile into a plateau or weak cusp
~\cite{Gondolo:1999ef,Vasiliev:2007vh,Shapiro:2016ypb}, 
while gravitational scattering by stars can soften the spike to a slope of $\alpha=3/2$
~\cite{Gnedin:2003rj,Merritt:2003qk,Merritt:2006mt,Shapiro:2022prq}. 
Furthermore, off-center black-hole formation 
and black-hole mergers can significantly weaken 
or even disrupt the spike~\cite{Ullio:2001fb,Merritt:2002vj}.

We adopt two benchmark spike profiles following Ref.~\cite{Cline:2023tkp}. 
The first, BM1, corresponds to adiabatic spike formation 
and is described by
$\rho_\chi(r) \propto(1-4R_S/r)^3(R_{\rm sp}/r)^{7/3}$.
The second, BM1$'$, 
incorporates stellar heating, 
which reduces the inner density slope to $\alpha=3/2$ 
within the black hole's radius of gravitational influence, 
$r_h=0.65$~pc for NGC~1068. 
The resulting profile scales as $\rho_\chi\propto(1-4R_S/r)^3r^{-3/2}$ for $r<r_h$ 
and is matched continuously to the outer $r^{-7/3}$ profile at $r_h$. 
In both benchmarks, the normalization is fixed by the condition 
$M_{\rm BH}=4\pi\int_{4R_S}^{r_h}r^2\rho_\chi(r)\,dr$, 
while $R_{\rm sp}$ is determined by matching the spike to the NFW host halo; other normalizations  (e.g., Ref.~\cite{Gondolo:1999ef}) yield smaller columns. We neglect $\chi_1$ annihilation, since, in the parameter region of interest, 
no two-body annihilation channel is kinematically accessible 
at tree level in the model of Sec.~\ref{sec:model}, 
and the remaining processes are too suppressed 
to generate an appreciable annihilation plateau.

Following Ref.~\cite{Cline:2023tkp}, we assume that neutrinos are emitted 
at a radius $R_{\rm em}$ and propagate radially outward, such that $d\ell=dr$. 
The DM column density along their trajectory is then 
$\Sigma=\int_{R_{\rm em}}^{100\,{\rm pc}}\rho_\chi(r)\,dr$, 
which is dominated by the region near the emission radius owing to the steep density profile. 
For the NGC~1068 corona, we adopt $R_{\rm em}=30R_S$ as our benchmark. 
The resulting column densities are summarized in Table~\ref{a:tab:sigma}. 
At this emission radius, the BM1 and BM1$'$ columns differ 
by a factor of approximately 650. Increasing $R_{\rm em}$ 
from $30R_S$ to $100R_S$ reduces $\Sigma$ by a factor of 4.2 and 1.7 for BM1 and BM1$'$, respectively. 
Results for other sources and further details of the spike modeling 
are provided in Appendix~\ref{app:multi}.

For stationary DM targets, the dependence on the spike profile enters 
the neutrino transport only through the asymptotic optical depth 
$\tau_0=\sigma_0\Sigma/m_1$. The column density therefore determines 
the relative optical depths in the multi-source analysis (Sec.~\ref{sec:other}) 
and enables the conversion of constraints on $\tau_0$ into bounds on 
the scattering cross section $\sigma_0$ 
and the underlying particle-physics parameters (see Sec.~\ref{sec:model}). 
The couplings responsible for the scattering also induce $\chi_1$ self-scattering, 
which can relax the inner spike and reduce the column.  We require the spike to remain collisionless when converting optical-depth limits into coupling limits in Sec.~\ref{sec:model}.

\begin{table}[t]
\centering
\begin{tabular}{lccc}
\hline\hline
$\Sigma$ [GeV\,cm$^{-2}$] & $R_{\rm em}=10R_S$ & $30R_S$ & $100R_S$\\
\hline
BM1 ($r^{-7/3}$) & $1.49\times10^{32}$ & $5.62\times10^{31}$ & $1.33\times10^{31}$\\
BM1$'$ ($r^{-3/2}$) & $1.18\times10^{29}$ & $8.70\times10^{28}$ & $5.21\times10^{28}$\\
\hline\hline
\end{tabular}
\caption{DM column densities $\Sigma$ for the non-annihilating spike 
benchmarks BM1 and BM1$'$ around the central supermassive black hole 
of NGC~1068 ($M_{\rm BH}=10^7M_\odot$), integrated from the neutrino emission 
radius $R_{\rm em}$ to 100~pc.}
\label{a:tab:sigma}
\end{table}

\section{NGC~1068 analysis} 
\label{sec:ngc}

IceCube track samples are primarily sensitive to the $\nu_\mu+\bar\nu_\mu$ flux at Earth, 
for which the publicly available effective areas are provided~\cite{Abbasi:2026ehs}. 
For the flavor-democratic neutrino flux assumed in our analysis, 
the muon-flavor component accounts for one third of the total transmitted flux $\Phi^{\rm out}$. 
The flux observed at Earth also incorporates geometric dilution and cosmological redshift, 
with neutrinos emitted at energy $E_{\nu,\,\rm source}=(1+z)E_\nu$ 
arriving with energy $E_\nu$. For NGC~1068, the redshift is sufficiently small that 
$E_{\nu,\,\rm source}\simeq E_\nu$. 
The observed muon-neutrino and antineutrino flux can therefore be expressed as
\begin{equation}
\Phi_\oplus(E_\nu;\tau_0,\Eth,\epsilon)
=\Phi_\oplus^{\rm in}(E_\nu)\,
R(E_\nu;\tau_0,\Eth,\epsilon),
\label{eq:Phiobs}
\end{equation}
where $\Phi_\oplus^{\rm in}$ denotes the $\nu_\mu+\bar\nu_\mu$ 
flux that would be observed at Earth in the absence of neutrino--DM scattering. 
We model this intrinsic (unattenuated) flux as a power law,
\begin{equation}
\Phi_\oplus^{\rm in}(E_\nu)
=\Phi_0\,(E_\nu/{\rm TeV})^{-\gamma},
\label{eq:PL}
\end{equation}
where $\Phi_0$ is the flux normalization at 1~TeV and $\gamma$ is the spectral index. 
Equations~\eqref{eq:Phiobs} and \eqref{eq:PL} define the common flux parametrization 
adopted in the two analyses below. 
For the event-count analysis, we fix $\gamma=3.2$, corresponding 
to the IceCube best-fit spectral index under the no-scattering hypothesis. 
In the likelihood analysis, in contrast, $\gamma$ is treated as a nuisance parameter 
and profiled jointly with the flux normalization at each point in the attenuation-parameter space.

\subsection{Event-count analysis}
\label{sec:eventcount}

As in Ref.~\cite{Cline:2023tkp}, we fix the intrinsic spectral index to the IceCube best-fit value without scattering, $\gamma=3.2$, 
and compute the expected number of track events with true neutrino energies 1.5--15~TeV,
\begin{equation}
\begin{aligned}
N(\tau_0,\Eth,\epsilon)=T_{\rm live}
\int_{1.5\,{\rm TeV}}^{15\,{\rm TeV}}
&dE_\nu\,A_{\rm eff}(E_\nu,\delta_{\rm s})\\
&\times\Phi_\oplus^{\rm in}(E_\nu)
R(E_\nu;\tau_0,\Eth,\epsilon),
\end{aligned}
\label{eq:N}
\end{equation}
where $T_{\rm live}$ is the live time and $A_{\rm eff}(E_\nu,\delta_{\rm s})$ is the IceTracks-DR2 IC86 effective area~\cite{Abbasi:2026ehs} 
at the source declination $\delta_{\rm s}$. Without scattering, $N_0\equiv N(0,\Eth,\epsilon)=31\pm7$~\cite{Cline:2023tkp}; 
requiring $N\ge22.2$, the 90\% confidence level (CL) lower bound adopted there, gives $N/N_0\ge0.716$, 
independent of $T_{\rm live}$ and the flux normalization. Pure absorption, $R=e^{-\tau_0}$, would give $\tau_0<0.33$, 
whereas elastic light-mediator scattering gives only $\tau_0<66$ for the benchmark of Ref.~\cite{Farzan:2026wek} ($m_{\rm DM}=1$~GeV, $m_{\rm med}^2/m_{\rm DM}=2$~GeV) and $\tau_0<7.9$ for $m_{\rm DM}=300$~keV and $m_{\rm med}=0.1$~GeV.

Figure~\ref{fig:constraints} shows the inelastic limits. For $\epsilon=0.01$ and $0.1~{\rm TeV}\le\Eth\lesssim2$~TeV, 
we obtain $\tau_0<1.1$--3.8, 3--11 times weaker than pure absorption and strongest near $\Eth\simeq1$~TeV. 
Since every scattering removes at least $\Eth$ and only neutrinos above $\Eth$ scatter, three regimes appear.
(i) For $\Eth\ll1$~TeV, scattered neutrinos stay in the counting window and are replaced by neutrinos from slightly higher energies, 
so $\Delta\simeq-\gamma\Eth/E_\nu$ and the limit weakens roughly as $1/\Eth$ ($\tau_0<3.8$ at 0.1~TeV).
(ii) For $\Eth\simeq1$~TeV, just below the window, each scattering moves most counted neutrinos out of the window, 
and regeneration from higher energies compensates only about half of the loss, giving the strongest limit, $\tau_0<1.1$.
(iii) For $\Eth$ inside the window, the spike can deplete only the events above $\Eth$, while those below it are unaffected or even enhanced by the pile-up. 
Since most counted events lie at a few TeV (median 3.3~TeV), the limit weakens rapidly ($\tau_0<5$ at 3~TeV, 14 at 3.7~TeV) 
and disappears at $\Eth\simeq4$~TeV; the pile-up never violates the upper bound $N\lesssim40$.
For $\epsilon=0.36$ (dashed blue curve), the limits are flatter, $\tau_0<1.15$--2.4 for $\Eth=0.1$--1.3~TeV. 
They are stronger than for $\epsilon=0.01$ below $\Eth\simeq0.6$~TeV ($\tau_0<2.4$ instead of 3.8 at 0.1~TeV), since regeneration is less effective, 
and weaker above it ($\tau_0<3.4$ instead of 2.1 at 2~TeV, and 19 instead of 6.0 at 3.2~TeV), since the broadened edge transmits more flux inside the window.

This criterion is an illustrative comparison with previous studies rather than a self-consistent analysis of the IceCube data. 
It fixes the intrinsic spectrum to the no-scattering best fit, although a larger normalization or a different index can compensate for much of the attenuation 
(pure absorption is exactly degenerate with the normalization); it counts events in a window of \emph{true} energy, 
whereas reconstructed energies are uncertain by $\gtrsim0.5$~dex (a factor of $\simeq3$; 1~dex is a factor of 10); it takes $N_0$ from Ref.~\cite{Cline:2023tkp}, which used a different effective area, and this choice matters: the pure-absorption limit $\tau_0<-\ln(22.2/N_0)$ rises from 0.33 at $N_0=31$ to 0.59, 0.81, and 0.99 for $N_0=40$, 50, and 60;
and it discards the energy and direction of individual events. 
The likelihood analysis of Sec.~\ref{sec:likelihood} removes these limitations and yields weaker but self-consistent limits.

\subsection{Likelihood analysis}\label{sec:likelihood}

We analyze the publicly available IceTracks-DR2 
dataset (14 seasons, 2008--2022)~\cite{Abbasi:2026ehs} 
using the unbinned point-source likelihood implemented 
in SkyLLH~\cite{Wolf:2019cfm,IceCube:2021mzg,IceCube:2023ihk}, 
the public framework employed for the DR2 benchmark fits. 
We parametrize the observed neutrino flux as in Eqs.~\eqref{eq:Phiobs} and \eqref{eq:PL}. 
The intrinsic flux normalization $\Phi_0$ 
and spectral index $\gamma$ are treated as nuisance parameters 
and profiled at each fixed point $(\tau_0,\Eth,\epsilon)$ in the attenuation-parameter space. 
We refer to this observed neutrino flux as the threshold-induced spectrum, 
which reduces to a pure power law in the absence of neutrino--DM scattering ($\tau_0=0$).

To account for the threshold-induced spectral feature, 
we introduce two modifications to the standard SkyLLH implementation. 
First, since the default framework supports 
only power-law and exponentially cutoff power-law spectra, 
we implement the threshold-induced spectrum of Eq.~\eqref{eq:Phiobs} as a custom SkyLLH energy profile:
the power law of Eq.~\eqref{eq:PL} multiplied by the transmission $R(E_\nu)$ from the full cascade,
tabulated in $\tau_0$ and $\gamma$ (see Appendix~\ref{app:transport}).
The spectral integrals over SkyLLH's coarse true-energy bins 
are evaluated numerically to controlled precision, 
ensuring that the threshold feature is properly captured 
even when it lies within a single bin. 
As a validation check, our implementation reproduces 
the standard SkyLLH power-law fit to numerical precision in the limit $\tau_0=0$.
Second, we disable the default three-bin smoothing applied 
to the data-driven background energy distribution. 
Although this smoothing has little impact on the standard power-law fits, 
we find that it alters the background distribution by up to $\pm15\%$ 
in the steeply falling high-energy tail relevant to the threshold feature. 
Pseudo-experiment tests reveal that retaining the smoothing leads to 
substantial undercoverage of the low-$\Eth$ upper limits, 
with an empirical coverage of only 61\% compared with the nominal 90\% CL 
in 150 pseudo-experiments.
Disabling the smoothing improves the coverage to 93\% for the tested configuration. 
We emphasize that these tests cover only a few thresholds and a single true spectrum; 
coverage elsewhere in the parameter space is not established.

The unbinned point-source likelihood of the DR2 analysis~\cite{Abbasi:2026ehs} 
is constructed as a product over the four DR2 data samples $d$ (IC40, IC59, IC79, and IC86-I--XI):
\begin{equation}
\mathcal L(n_s,\gamma;\Eth,\epsilon,\tau_0)=
\prod_d\prod_{i\in d}
\left[
\frac{n_sf_d}{N_d}\mathcal S_i+
\left(1-\frac{n_sf_d}{N_d}\right)\mathcal B_i
\right],
\label{eq:like}
\end{equation}
where $N_d$ is the number of events in sample $d$, 
$f_d$ is the fraction of the total expected signal events $n_s$ 
contributed by that sample ($\sum_df_d=1$), and $\mathcal S_i$ ($\mathcal B_i$) 
denotes the signal (background) probability density 
for event $i$ in reconstructed direction and energy. 
All events of the four samples enter, over the full reconstructed-energy range of the release.  The signal energy PDFs are built from the instrument response functions over true neutrino energies $10^2$--$10^9$~GeV~\cite{Abbasi:2026ehs}. 
The spectral parameters $\gamma$, $\Eth$, $\epsilon$,
and $\tau_0$ enter through $\mathcal S_i$ and $f_d$, 
while $n_s$ determines the overall signal normalization 
and is equivalent to $\Phi_0$. The kernel width $\epsilon$ is not a fit parameter: each scan is performed at a fixed $\epsilon$, here one of the two benchmark values 0.01 and 0.36. The test statistic is
\begin{equation}
{\rm TS}(\Eth, \epsilon, \tau_0)=
-2\ln\frac{\mathcal L(n_s=0)}
{\mathcal L(\hat n_s,\hat\gamma;\Eth,\epsilon,\tau_0)}
\ge0,
\label{eq:TS}
\end{equation}
where $\hat n_s\ge0$ and $\hat\gamma\in[2,4]$ (extended to $\hat\gamma\ge1.5$ for TXS~0506+056 in Sec.~\ref{sec:other}) maximize
the likelihood at fixed $\Eth$, $\epsilon$, and $\tau_0$. 
The case $\tau_0=0$ reduces to the standard power-law hypothesis.

For each fixed $\Eth$ and $\epsilon$, we determine the best-fit optical depth $\hat\tau_0(\Eth,\epsilon)$ 
by maximizing ${\rm TS}$ and construct the profile-likelihood difference
\begin{equation}
\Delta{\rm TS}(\Eth,\epsilon,\tau_0)=
{\rm TS}(\Eth,\epsilon,\hat\tau_0)-{\rm TS}(\Eth,\epsilon,\tau_0).
\label{eq:dTS}
\end{equation}
We define, for each fixed $\epsilon$, the upper limit $\tau_{0,90}(\Eth)$
as the first value of $\tau_0>\hat\tau_0$ 
satisfying $\Delta{\rm TS}=2.71$ 
(for all curves shown, $\Delta{\rm TS}$ remains above 2.71 
beyond this first crossing, so the excluded region is $\tau_0>\tau_{0,90}$), 
retaining the nominal 90\% CL convention used in our figures. 
This prescription follows the likelihood-ratio approach: 
in the regular one-parameter Gaussian limit, $\Delta{\rm TS}=2.71$ 
corresponds to a central 90\% two-sided confidence interval, 
whose upper endpoint is equivalent to a 95\% one-sided bound. 
This correspondence can be modified by parameter boundaries, 
nuisance parameters, and finite event statistics. 

To quantify the preference for a threshold-induced 
spectral feature over a pure power law, we further define
\begin{equation}
\Delta{\rm TS}_{\rm th}(\Eth,\epsilon)=
{\rm TS}(\Eth,\epsilon,\hat\tau_0)-{\rm TS}(\Eth,\epsilon,0).
\label{eq:dTSstep}
\end{equation}
By construction, $\Delta{\rm TS}_{\rm th}\ge0$, with $\Delta{\rm TS}_{\rm th}=0$ when $\hat\tau_0=0$. 
The resulting limits are evaluated pointwise in $\Eth$, 
and we do not assign a global significance to the threshold scan.

In the absence of DM attenuation, 
the best-fit power-law spectrum yields $\hat\gamma=3.10$, $\hat n_s=72.5$ signal events, 
and ${\rm TS}=25.4$, consistent with the published DR2 results for NGC~1068~\cite{Abbasi:2026ehs}. 

\begin{figure}[tbp]\centering
\includegraphics[width=\columnwidth]{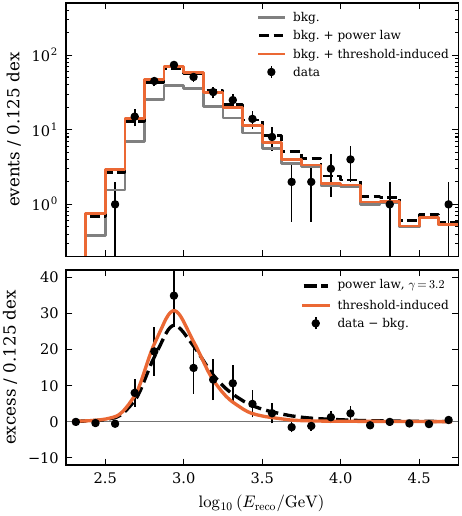}
\caption{
Top: IceTracks-DR2 events within $1^\circ$ of NGC~1068 (14 seasons) per 0.125-dex bin of reconstructed energy, background estimated from the
3--$10^\circ$ ring (bkg.), and background plus signal for a power law and for the threshold-induced spectrum at $\Eth=1.67$~TeV ($\tau_0=50$, $\epsilon=0.01$).
Bottom: excess over background and the signal shapes (SkyLLH, $\gamma=3.2$; responses of IC40, IC59, IC79, and IC86-I--XI, weighted by their expected signal events), each normalized to the total excess.}
\label{fig:data}
\end{figure}

Figure~\ref{fig:data} compares the observed event distribution 
near the source with the expected signal. 
The excess of $105\pm17$ events within $1^\circ$ concentrates
at reconstructed energies of 0.5--2.5~TeV, 
whereas the predictions of a pure power law 
and a spectrum with an edge at $\Eth=1.67$~TeV 
differ primarily in the sparsely populated high-energy tail above $\approx2.5$~TeV. 
Because the reconstructed energy resolves the neutrino energy only to $\gtrsim0.5$~dex, the edge is smeared over several bins: 
the threshold-induced spectrum appears as a deficit of signal at reconstructed high energies and, once normalized to the observed excess, 
as a mild shift of the signal toward lower energies, rather than as a sharp feature. 
With $\gamma=3.2$ fixed and each shape normalized to the excess, a simple binned comparison of the 20 bins 
reveals no statistically significant difference in the Poisson deviance between the pure power-law spectrum and the threshold-induced spectrum at $\tau_0=50$. 
The discriminating power therefore rests on the few events above $\approx2.5$~TeV, 
and quantifying it requires the unbinned likelihood with directional information and a profiled spectral index described above. 

\begin{figure}[tbp]\centering
\includegraphics[width=\columnwidth]{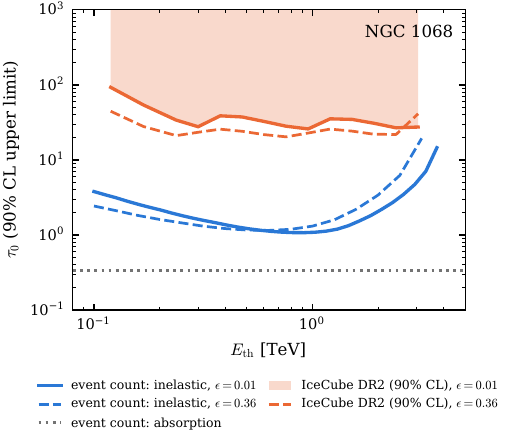}
\caption{Upper limits on the asymptotic optical depth $\tau_0$ as a function of the inelastic threshold $\Eth$ for NGC~1068. The shaded region (orange) indicates the pointwise exclusion derived from the IceTracks-DR2 likelihood analysis (nominal 90\% CL, $\Delta{\rm TS}=2.71$). The solid blue curve shows the limit obtained from the event-count criterion, while the dotted grey line represents the corresponding pure-absorption limit, neglecting neutrino regeneration. Solid curves and shading: $\epsilon=0.01$; dashed curves: $\epsilon=0.36$.}
\label{fig:constraints}
\end{figure}

The constraints on $\tau_0$ as a function of $\Eth$ 
derived from the NGC~1068 IceTracks-DR2 data 
are shown in Fig.~\ref{fig:constraints}. 
For the narrow kernel, $\epsilon=0.01$, 
we obtain upper limits of $\tau_0<26$--39 for $\Eth=0.24$--3.0~TeV, 
which weaken to $\tau_0<54$ and 93 at $\Eth=0.17$ and 0.12~TeV, respectively (solid orange curve). 
No upper limit is obtained for $\Eth\gtrsim3.5$~TeV. 
For the broad kernel, $\epsilon=0.36$, we obtain $\tau_0<20$--26 for $\Eth=0.24$--2.4~TeV, 
and $\tau_0<28$ and 45 at $\Eth=0.17$ and 0.12~TeV, respectively, 
whereas at $\Eth=3.0$~TeV the limit weakens to $\tau_0<41$ 
(dashed orange curve). 
Overall, the data show no significant preference for a threshold-induced spectral feature, 
with $\Delta{\rm TS}_{\rm th}\le0.19$ throughout the range $\Eth=0.12$--4.8~TeV.  
The dependence on the kernel width is discussed in Appendix~\ref{app:kernelwidth}.

These limits assume a pure power-law spectrum for $\Phi_\oplus^{\rm in}$. 
Because only the product $\Phi_\oplus^{\rm in}R$ is constrained, 
additional freedom in the intrinsic shape can absorb part of the scattering-induced distortion, 
as the difference between the event-count and likelihood limits already illustrates 
for the normalization and index. 
The most relevant extension is a high-energy cutoff: 
the threshold-induced spectrum differs from a power law not only near $\Eth$ 
but also far above it, where the transmission recovers (i.e., $R \simeq 1$), 
and the intrinsic spectrum of NGC~1068 above the energies probed by the data is not constrained.
To assess this dependence for both kernels, we multiply the power-law parametrization in Eq.~\eqref{eq:PL} 
by an exponential cutoff, $e^{-E_\nu/E_c}$, and refer to the resulting observed spectrum as the cutoff threshold-induced spectrum. We profile over eleven values $E_c=0.5$--500~TeV, together with the case without a cutoff ($E_c=\infty$),
and over $\gamma\in[2,4]$. In the absence of DM scattering, 
the cutoff provides no significant improvement in the fit, 
increasing the test statistic by only 0.12.
Nevertheless, allowing a cutoff substantially weakens the attenuation constraints, 
yielding $\tau_0<61$--134 for $\Eth=0.3$--3.0~TeV and $\tau_0<692$ at $\Eth=0.12$~TeV for the narrow kernel.
For the broad kernel, $\epsilon=0.36$, we repeat the scan at six thresholds, $\Eth=0.24$--2.4~TeV, with the same cutoff grid.
Without a cutoff, it reproduces the baseline $\epsilon=0.36$ limits to within 0.5\%.
Allowing a cutoff weakens the limits by similar factors, to $\tau_0<44$--105, 
i.e., by factors of 5.0 and 3.7 at $\Eth=0.24$ and 0.30~TeV,
3.3--3.6 at 0.6--1.0~TeV, and 2.8 and 2.0 at 1.5 and 2.4~TeV.
The weakest constraints typically arise for cutoff energies $E_c\sim10$--100~TeV,
well above the threshold-induced spectral feature.

This behavior can be understood from the high-energy recovery of the transmitted spectrum. 
At $E_\nu\gg\Eth$, the fractional energy loss per scattering becomes small, 
allowing regeneration to increasingly compensate for depletion. 
For example, at $\tau_0=32$ and $\gamma=3$, 
the transmission increases from $R=0.12$ at $E_\nu=30\Eth$ to $R=0.43$ at $E_\nu=100\Eth$. 
An exponential cutoff can preserve part of this recovered flux at energies of order $10$--$100\Eth$, 
which contributes to the TeV signal through the detector response, 
while suppressing the higher-energy flux that is disfavored by the data. 
This interplay allows the cutoff threshold-induced spectrum to accommodate larger optical depths. 
Indeed, artificially suppressing the high-energy recovery reduces 
the weakening of the limits to at most a factor of 2.4, 
demonstrating its importance. 
For $\Eth\gtrsim0.6$~TeV, imposing the additional restriction 
$\gamma\ge2.5$ also affects the constraints appreciably, 
since the weakest limits in this region favor hard intrinsic spectra with $\gamma\simeq2$.

We emphasize that the cutoff-profiled limits exhibit numerical sensitivity to the sampling of $E_c$, 
changing by up to 30\% upon refinement of the cutoff-energy grid. 
They should therefore be regarded as estimates of the potential impact of intrinsic spectral uncertainties 
rather than precisely calibrated bounds. 
We retain the pure power-law spectrum as our baseline, 
since the data show no statistically significant preference for an exponential cutoff. 

\section{Other sources}\label{sec:other}

Neutrino sources with harder spectra can probe higher inelastic thresholds. 
We therefore extend our analysis to the Seyfert galaxy CGCG~420-015, 
for which IceCube has reported a neutrino excess alongside NGC~4151~\cite{IceCube:2024dou}, 
and the blazar TXS~0506+056~\cite{IceCube:2018cha,Abbasi:2026ehs}. 
The threshold in the observed spectrum is redshifted relative to its source-frame value, 
$E_{\rm th,obs}=E_{\rm th,src}/(1+z)$, 
with $z=0.030$~\cite{Inoue:2024nap} and 0.337~\cite{Paiano:2018qeq} for CGCG~420-015 and TXS~0506+056, respectively. 
We combine these sources with NGC~1068 to constrain a common particle-physics model, 
in which the relative optical depths are determined by the corresponding DM spike columns. 
For our benchmark spike prescriptions, the column density of CGCG~420-015, 
with $M_{\rm BH}=2\times10^8M_\odot$~\cite{Koss:2017ilk}, 
is approximately 5\% of that of NGC~1068, 
while the corresponding ratio for TXS~0506+056 
ranges from $10^{-4}$ to $5\times10^{-3}$ (see Appendix~\ref{app:multi}). 
We emphasize that the interpretation of the additional sources 
as neutrino emitters, and hence their contribution to the combined constraints, 
is conditional on the reported $\sim3\sigma$ excesses being of astrophysical origin. 
Here the multi-source analysis is performed for the narrow kernel only ($\epsilon=0.01$).

For each source $a$, we construct 
the unbinned point-source likelihood following Eqs.~\eqref{eq:like} and \eqref{eq:TS}. 
At a fixed source-frame threshold, the corresponding test statistic is
\begin{equation}
{\rm TS}_a(\Eth,\tau_a)=
-2\ln\frac{\mathcal L_a(n_{s,a}=0)}
{\mathcal L_a(\hat n_{s,a},\hat\gamma_a;\Eth,\tau_a)},
\label{eq:TSi}
\end{equation}
where $\mathcal L_a$ is the likelihood evaluated at the position of source $a$, 
with the threshold feature appearing at the redshifted energy $E_{\rm th,obs}$. 
Each source has an independent intrinsic power-law normalization $n_{s,a}$ 
and spectral index $\gamma_a$, which are profiled at fixed $\Eth$ and $\tau_a$.

For a common particle-physics model specified by $\sigma_0$, $m_1$, and $\Eth$, 
the optical depths differ only through the DM column densities,
$\tau_a=\kappa_a\tau_0$, where $\kappa_a\equiv\Sigma_a/\Sigma_{\rm NGC\,1068}$ 
and $\tau_0$ denotes the asymptotic optical depth of NGC~1068. 
Assuming statistically independent source likelihoods, the joint test statistic is
\begin{equation}
{\rm TS}_{\rm joint}(\Eth,\tau_0)
=\sum_a{\rm TS}_a(\Eth,\kappa_a\tau_0).
\label{eq:TSjoint}
\end{equation}
This approximation is supported by the angular separation of 
the sources: CGCG~420-015 and TXS~0506+056, the closest pair, 
are separated by $4.2^\circ$, substantially larger than the typical IceCube track 
angular resolution of $\lesssim1^\circ$ at TeV energies. 
The profile-likelihood difference $\Delta{\rm TS}$ 
and the corresponding upper limit $\tau_{0,90}$ are then obtained 
from Eq.~\eqref{eq:dTS}, replacing ${\rm TS}$ with ${\rm TS}_{\rm joint}$, 
as in the NGC~1068 analysis.

To assess the preference for a threshold-induced spectral feature 
without imposing the spike-column relations, 
we also allow the optical depth of each source to vary independently. 
Defining $\hat\tau_a(\Eth)$ as the value that maximizes ${\rm TS}_a$ at fixed $\Eth$, 
the improvement over a pure power law is
$\Delta{\rm TS}_{{\rm th},a}(\Eth)
={\rm TS}_a(\Eth,\hat\tau_a)-{\rm TS}_a(\Eth,0)$.
For a common source-frame threshold, 
the combined improvement with independently fitted optical depths is therefore $\sum_a\Delta{\rm TS}_{{\rm th},a}(\Eth)$.
This provides a complementary test of the spectral feature 
without assuming a common normalization of the DM scattering optical depths across sources.

We scan 22 source-frame thresholds spanning $\Eth=0.12$--78~TeV, with a spacing of 0.1~dex over 0.3--4.8~TeV and 0.15--0.2~dex outside this range. For each source, the likelihood is evaluated at 15 optical depths spanning $\tau_a=0$--$10^3$. We interpolate ${\rm TS}_a$ in $\ln\tau_a$ for $\tau_a>0$ and hold it fixed beyond $\tau_a=10^3$. To assess the impact of this extrapolation, we extend the fits to $\tau_a=10^5$. The test statistic changes by at most 0.5 for NGC~1068 and 1.7 for CGCG~420-015, while the TS for TXS~0506+056 approaches zero by $\tau_a\simeq10^4$. None of the reported limits is affected, indicating that the adopted optical-depth range is sufficient for the constraints presented here.

A combined analysis of multiple IceCube sources was previously performed in Ref.~\cite{Dixit:2026zkv} 
to constrain the elastic neutrino--DM scattering cross section. 
Here, we extend this approach to inelastic scattering using an event-level likelihood 
with independent intrinsic spectra for each source. 
The source properties, DM column densities, and corresponding optical-depth ratios $\kappa_a$ 
are summarized in Table~\ref{a:tab:sources}.

\begin{figure}[tbp]\centering
\includegraphics[width=\columnwidth]{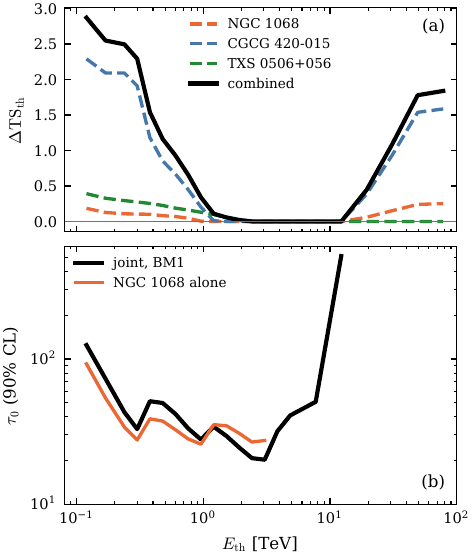}
\caption{Multi-source analysis (SkyLLH baseline, source-frame thresholds, $\epsilon=0.01$). 
(a) TS improvement of the threshold-induced spectrum over the power-law spectrum,
$\Delta{\rm TS}_{{\rm th},a}$ for each source (dashed) and combined $\Delta{\rm TS}_{\rm th}$ with free depths
(solid). (b) Joint nominal 90\% CL limit on $\tau_0$ (NGC~1068 normalization), with the depths tied to the BM1
spike columns (black), and the limit from NGC~1068 alone (orange).}
\label{fig:dTS}
\end{figure}

Using the IceTracks-DR2 data set with SkyLLH, 
the baseline power-law fits ($\tau_a = 0$) yield ${\rm TS}=12.5$ ($\hat n_s=42.7$, $\hat\gamma=2.68$) for CGCG~420-015 
and ${\rm TS}=11.8$ ($\hat n_s=10.0$, $\hat\gamma=1.96$) for TXS~0506+056.
For the latter, we extend the allowed spectral-index range and the transport tables to $\gamma\ge1.5$, 
since restricting $\gamma\ge2$ causes many fits with the threshold-induced spectrum to accumulate at the lower boundary. 

Figure~\ref{fig:dTS}(a) shows the improvement in the fit obtained 
by introducing a threshold-induced spectral feature. 
For each source, $\Delta{\rm TS}_{{\rm th},a}(\Eth)$ 
compares the best-fit threshold-induced spectrum, 
with $\tau_a$, $n_{s,a}$, and $\gamma_a$ allowed to vary, 
against the corresponding pure power-law fit. 
Since the power-law hypothesis is recovered at $\tau_a=0$, $\Delta{\rm TS}_{{\rm th},a}\ge0$, 
with zero indicating no improvement over the power law. 
Neither NGC~1068 nor TXS~0506+056 shows an appreciable 
preference for a threshold-induced feature, 
with $\Delta{\rm TS}_{{\rm th},a}\le0.25$ and 0.39, respectively, 
throughout the threshold range considered (for NGC~1068, the largest value, 0.25, occurs at $\Eth=49$--78~TeV; below 4.8~TeV it is $\le0.19$, as in Sec.~\ref{sec:likelihood}). 
In particular, all three sources yield $\Delta{\rm TS}_{{\rm th},a}=0$ for $\Eth=3$--12~TeV. 

Only CGCG~420-015 exhibits a mild preference, 
reaching $\Delta{\rm TS}_{{\rm th},a}=2.3$ at $\Eth=0.12$~TeV with $\hat\tau_a=46$, 
and approximately 1.6 for $\Eth\gtrsim50$~TeV. 
In the latter regime, the best-fit optical depth reaches $\hat\tau_a=10^3$, 
effectively suppressing the flux above the threshold and mimicking 
a high-energy cutoff of the intrinsic power-law spectrum with $\hat\gamma_a=2.68$. 
Allowing the optical depths to vary independently, 
the combined improvement at a common source-frame 
threshold reaches a maximum of $\Delta{\rm TS}_{\rm th}=2.9$ at $\Eth=0.12$~TeV. 
The same maximum is obtained when both the thresholds 
and optical depths are allowed to vary independently among sources.
This modest improvement arises after scanning 22 thresholds with up to three independent optical depths. 
We therefore do not assign a global significance to the excess, 
and emphasize that the upper-limit criterion $\Delta{\rm TS}=2.71$ 
does not constitute a discovery threshold for this scan.

When the optical depths are tied to the DM spike columns through $\tau_a=\kappa_a\tau_0$, 
the preference at $\Eth=0.12$~TeV decreases substantially, from $\Delta{\rm TS}_{\rm th}=2.9$ to 0.23, 
with a best-fit value $\hat\tau_0=7$ for both BM1 and BM1$'$. 
This reduction can be understood from the tension between the sources: 
the optical depth preferred by CGCG~420-015 alone, $\hat\tau_a=46$, 
corresponds to $\tau_0\simeq900$ for NGC~1068, 
well above its individual upper limit of $\tau_0<93$. 
For $\Eth<12$~TeV, the improvement in the tied-depth fit does not exceed 0.23. 
The largest improvement occurs at $\Eth=78$~TeV, 
reaching $\Delta{\rm TS}_{\rm th}=1.8$ (1.5) for BM1 (BM1$'$), 
driven by the cutoff-like feature in CGCG~420-015 discussed above. 
In this regime, however, the joint fit does not provide a meaningful constraint on the optical depth.

For the combined constraints, 
we adopt $\kappa_a=0.05$ for CGCG~420-015 and $\kappa_a=1.4\times10^{-4}$ ($4.5\times10^{-3}$) 
for TXS~0506+056 under the BM1 (BM1$'$) spike prescription. 
For BM1, the joint likelihood yields $\tau_0<20$--51 over $\Eth=0.3$--7.7~TeV, 
weakening to $\tau_0<43$--125 at $\Eth=0.12$--0.24~TeV 
and $\tau_0<510$ at $\Eth=12$~TeV, as shown in Fig.~\ref{fig:dTS}(b). 
The combined analysis thus extends the threshold reach 
from approximately 3~TeV for NGC~1068 alone to about 12~TeV. 
Above this threshold, the remaining sensitivity comes primarily from TXS~0506+056, 
corresponding to optical depths $\tau_0\gtrsim2\times10^6$ for BM1 ($\gtrsim6\times10^4$ for BM1$'$) in the NGC~1068 normalization
because of its much smaller column density.

At low thresholds, the joint limits can be weaker than those from NGC~1068 alone, 
reflecting the mild preference for a threshold-induced feature 
in CGCG~420-015 as shown in Fig.~\ref{fig:dTS}(a). 
For example, the upper limits increase from 93 to 125 at $\Eth=0.12$~TeV 
and from 28 to 33 at $\Eth=0.3$~TeV. In contrast, the combined constraints 
become stronger for $\Eth=1.5$--3~TeV, improving 
from $\tau_0<27$--35 for NGC~1068 alone to $\tau_0<20$--30. 
At the benchmark threshold $\Eth=1.67$~TeV, 
we obtain $\tau_0<27.4$ for both $R_{\rm em}=30R_S$ and $100R_S$. 
For BM1$'$, the joint limits agree with those of BM1 to within 5\% for $\Eth<8$~TeV, 
while the upper limit at 12~TeV becomes $\tau_0<360$. 
This near-independence arises because the relative column density of CGCG~420-015 
is similar in the two spike benchmarks. 
Consequently, the optical-depth constraint is relatively insensitive to the spike prescription, 
whereas its conversion into a scattering cross-section limit depends 
directly on the absolute column density of NGC~1068. 
The individual source constraints, expressed in terms of their respective optical depths, 
are given in Appendix~\ref{app:multi}.

Using $\tau_0=\sigma_0\Sigma_{\rm NGC\,1068}/m_1$ as in Eq.~\eqref{eq:tau}, the upper limit on the asymptotic scattering cross section is
\begin{equation}
\sigma_{0,90}(\Eth)=
\frac{m_1}{\Sigma_{\rm NGC\,1068}}\,
\tau_{0,90}(\Eth).
\label{eq:sigma90}
\end{equation}
For the BM1 benchmark with $R_{\rm em}=30R_S$, $\Sigma_{\rm NGC\,1068}=5.6\times10^{31}$~GeV\,cm$^{-2}$, and $\epsilon=0.01$, the combined data constrain $\sigma_0/m_1<(3.6$--$9.1)\times10^{-31}$~cm$^2$\,GeV$^{-1}$ over $\Eth=0.3$--7.7~TeV. For $m_1=300$~keV, this corresponds to $\sigma_0<(1.1$--$2.7)\times10^{-34}$~cm$^2$, weakening to $6.7\times10^{-34}$~cm$^2$ at $\Eth=0.12$~TeV and $2.7\times10^{-33}$~cm$^2$ at $\Eth=12$~TeV. Above 12~TeV, only substantially weaker constraints from TXS~0506+056 remain within the scanned parameter range. For BM1$'$, the NGC~1068 column density is approximately 650 times smaller, leading to correspondingly weaker cross-section limits, $\sigma_0<(0.7$--$1.8)\times10^{-31}$~cm$^2$ for $\Eth=0.3$--7.7~TeV.

These constraints are conditional on the assumed inelastic scattering kernel, prompt-decay secondary-neutrino spectrum, intrinsic source spectra, and DM spike columns. Their interpretation in terms of particle-physics parameters and comparison with laboratory constraints are discussed in Sec.~\ref{sec:model}.

\section{Particle-physics interpretation}
\label{sec:model}

A concrete realization of the phenomenological scattering kernel in 
Eq.~\eqref{eq:sigma} is provided by a dark sector containing two Majorana fermions, 
$\chi_1$ and $\chi_2$, coupled through a vector mediator $Z'$ (see Appendix~\ref{app:model}). 
The relevant interactions are
\begin{equation}
\mathcal L\supset
ig_\chi Z'_\mu\bar\chi_2\gamma^\mu\chi_1
+g_\nu Z'_\mu\bar\nu\gamma^\mu P_L\nu,
\label{eq:L}
\end{equation}
where $g_\chi$ and $g_\nu$ denote the dark-sector and neutrino couplings, respectively, 
and $P_L$ is the left-handed chiral projector. 
We treat $g_\nu$ as an effective coupling that is flavor universal and diagonal, 
ensuring identical scattering interactions for all neutrino flavors and allowing 
the transport to be treated independently of flavor oscillations. 
We note that a coupling of $Z'$ exclusively to the left-handed neutrino current is not invariant 
under the Standard Model $SU(2)_L$ gauge symmetry and therefore requires an ultraviolet completion, 
which lies beyond the scope of this work.

Averaging over the initial $\chi_1$ spin, the squared matrix element for $\nu\chi_1\to\nu\chi_2$ is
\begin{equation}
\begin{split}
\overline{|\mathcal M|^2}
=&
\frac{2g_\nu^2g_\chi^2}{(t-m_{Z'}^2)^2}
\big[(s-m_1^2)(s-m_2^2)\\
& \qquad
+(u-m_1^2)(u-m_2^2) +2m_1m_2t\big],
\end{split}
\label{eq:msq}
\end{equation}
where $s$, $t$, and $u$ are the Mandelstam variables. 
In the regime $(E_\nu-\Eth)/\Eth\gg\epsilon$, with $\epsilon=m_{Z'}^2/(m_2^2-m_1^2)$, 
the exact cross section derived in Appendix~\ref{app:model} 
approaches the phenomenological form of Eq.~\eqref{eq:sigma}, 
with the asymptotic normalization
\begin{equation}
\sigma_0=\frac{g_\nu^2g_\chi^2}{4\pi m_{Z'}^2}.
\label{eq:map}
\end{equation}
In the hierarchical mass regime $m_1\ll m_2$ considered here, 
the narrow-kernel condition $\epsilon\ll1$ corresponds to $m_{Z'}\ll m_2$. 
Physically, the light mediator favors small momentum transfers, 
concentrating the neutrino energy loss near its kinematic minimum $\Eth$. 
The broad-kernel benchmark $\epsilon=0.36$ corresponds to $m_{Z'}\simeq0.6\,(m_2-m_1)$, 
for which the decay $\chi_2\to\chi_1Z'$ remains open, and our treatment of the secondary neutrinos applies without change.  The effect of the kernel width on the limits is shown in Fig.~\ref{a:fig:epsscan} of Appendix~\ref{app:kernelwidth}.

As the mediator mass increases, the energy-transfer distribution broadens, 
shifting the transmission minimum toward higher energies (see Fig.~\ref{fig:f}(c)). 
Moreover, when $m_{Z'}>m_2-m_1$, the two-body decay $\chi_2\to\chi_1Z'$ 
becomes kinematically forbidden, 
and the on-shell decay description underlying our secondary-neutrino treatment no longer applies.

Using $m_{Z'}^2=2\epsilon m_1\Eth$ and Eq.~\eqref{eq:map}, the asymptotic optical depth can be expressed in terms of the model parameters as
\begin{equation}
\tau_0=\frac{g_\nu^2g_\chi^2\,\Sigma}{8\pi\epsilon\,\Eth m_1^2},
\label{eq:tau_model}
\end{equation}
where consistent natural units are understood. For the benchmark parameters $\Eth=1.67$~TeV, $\epsilon=0.01$, and the BM1 column density $\Sigma=5.6\times10^{31}$~GeV\,cm$^{-2}$, we obtain
\begin{equation}
\begin{aligned}
\tau_0\simeq 33\,
&\left(\frac{g_\nu g_\chi}{2.4\times10^{-4}}\right)^2
\left(\frac{300~{\rm keV}}{m_1}\right)^2 \left(\frac{0.01}{\epsilon}\right)\\
&\times
\left(\frac{1.67~{\rm TeV}}{\Eth}\right)
\left(\frac{\Sigma}{5.6\times10^{31}~{\rm GeV\,cm}^{-2}}\right).
\end{aligned}
\label{eq:tau_benchmark}
\end{equation}
The reference normalization corresponds to the NGC~1068 likelihood limit $\tau_{0,90}\simeq33$ at this threshold, yielding $g_\nu g_\chi<2.4\times10^{-4}$ for the BM1 spike. An optical depth of order unity is reached for $g_\nu g_\chi\simeq4\times10^{-5}$, illustrating that sub-MeV DM in a dense spike can produce appreciable neutrino attenuation even at TeV-scale inelastic thresholds. Translating the optical-depth constraints into bounds on the underlying couplings, however, requires specifying the DM column density.

\begin{figure}[tbp]\centering
\includegraphics[width=\columnwidth]{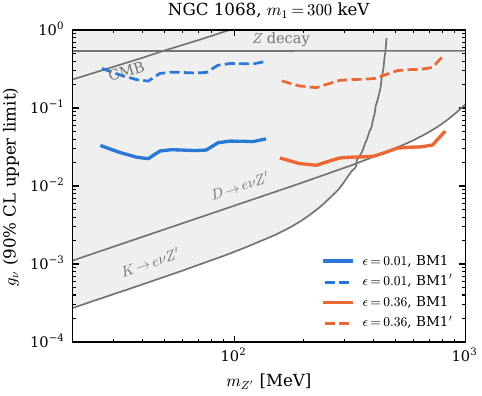}
\caption{
Nominal 90\% CL ($\Delta{\rm TS}=2.71$) limits on $g_\nu$ from the NGC~1068 likelihood analysis for $m_1=300$~keV, as a function of $m_{Z'}=(2\epsilon m_1\Eth)^{1/2}$
for $\epsilon=0.01$ (blue) and 0.36 (orange), with $\Eth=0.12$--3.0~TeV, in the dark-Higgs completion model. 
The coupling $g_\chi$ is set to the largest value for which the DM spike remains collisionless ($y=y_c$; see text).
Solid: adiabatic spike (BM1, $R_{\rm em}=30R_S$); dashed: stellar-heated spike (BM1$'$). 
Grey: laboratory and cosmological bounds from Ref.~\cite{Dev:2025tdv}: meson decays from its Fig.~7(a), and invisible $Z$ decays and the cosmic microwave background (CMB) from its Fig.~10; shaded regions are excluded.}
\label{fig:gnu}
\end{figure}

Figure~\ref{fig:gnu} shows the corresponding limits on the neutrino coupling $g_\nu$ as a function of the mediator mass $m_{Z'}$, derived from the NGC~1068 likelihood analysis of Sec.~\ref{sec:likelihood}. We fix $m_1=300$~keV and consider two kernel widths, $\epsilon=0.01$ and 0.36. At fixed $\epsilon$, the mediator mass and inelastic threshold are related by $m_{Z'}^2=2\epsilon m_1\Eth$. Thus, the threshold range $\Eth=0.12$--3.0~TeV corresponds to $m_{Z'}=27$--135~MeV for $\epsilon=0.01$ and $m_{Z'}=0.16$--0.81~GeV for $\epsilon=0.36$.

An important consideration in this interpretation is the self-consistency of the assumed DM spike. Throughout our transport analysis, we adopt the collisionless spike profiles BM1 and BM1$'$ of Sec.~\ref{sec:spike}, assuming that DM self-interactions do not significantly alter their density distributions. In the dark-Higgs completion (Appendix~\ref{app:model}), however, the dark scalar $h_D$ mediates $\chi_1\chi_1$ self-scattering at tree level, with an interaction strength controlled by the Yukawa coupling $y\equiv g_\chi(m_2-m_1)/m_{Z'}$. If self-scattering becomes sufficiently efficient, the inner spike enters a collisional regime and can relax toward a shallower density profile, reducing the DM column density and consequently the optical depth at fixed scattering couplings.

To ensure consistency with the collisionless spike assumption, we require the expected number of self-scatterings per $\chi_1$ particle at the emission radius $R_{\rm em}$ over the black-hole lifetime $t_{\rm BH}$ to be smaller than unity. Adopting the minimum tree-level self-scattering rate compatible with perturbative unitarity, this condition imposes an upper bound $y<y_c$, where $y_c$ is determined in Eq.~\eqref{a:eq:yc} of Appendix~\ref{app:selfconsistent}. For $t_{\rm BH}=10^8$~yr and $\Eth=0.12$--3.0~TeV, we find $y_c=0.033$--0.074 for BM1 and $y_c=0.084$--0.19 for BM1$'$. The critical coupling scales as $y_c\propto(m_2-m_1)^{1/2}$ and depends only weakly on the spike properties, with $y_c\propto(\rho_\chi t_{\rm BH})^{-1/8}$.

To obtain the strongest coupling constraints consistent with a collisionless spike, we set $y=y_c$, corresponding to the largest allowed dark-sector coupling. This determines $g_\chi=y_c m_{Z'}/(m_2-m_1)$ at each threshold. For BM1, we obtain $g_\chi=0.0033$--0.0074 and 0.020--0.044 for $\epsilon=0.01$ and 0.36, respectively. Combining Eqs.~\eqref{eq:tau} and \eqref{eq:map}, the corresponding upper limit on the neutrino coupling becomes
\begin{equation}
g_{\nu,90}=
\frac{m_2-m_1}{y_c}
\left[
\frac{4\pi m_1\tau_{0,90}}{\Sigma}
\right]^{1/2},
\label{eq:gnu_c}
\end{equation}
as shown in Fig.~\ref{fig:gnu} for BM1 (solid) and BM1$'$ (dashed). Smaller values of $y$ preserve the collisionless spike but weaken the constraint on $g_\nu$ proportionally to $1/y$.

For BM1, we obtain $g_\nu<0.018$--0.049 for the two benchmark kernels. 
The kernel width nevertheless determines the mediator mass range over which these limits apply. 
For the narrow kernel ($\epsilon=0.01$), the resulting constraints are weaker than the laboratory bounds from $K\to e\nu Z'$ and $D^+\to e^+\nu Z'$ decays~\cite{Dev:2025tdv}.\footnote{Since ${\rm BR}(Z'\to\nu\bar\nu)=1$ in the parameter region of interest, the meson-decay bounds constrain $g_\nu$ alone, and we adopt the $K^+\to e^+\nu Z'$ and $D^+\to e^+\nu Z'$ bounds computed for $M^\pm\to\ell^\pm\nu Z'$ in Fig.~7(a) of Ref.~\cite{Dev:2025tdv}.} 
For the broader kernel ($\epsilon=0.36$), the laboratory bounds remain stronger at $m_{Z'}\lesssim0.45$~GeV, whereas the NGC~1068 constraints become more stringent for larger $m_{Z'}$. Allowing an intrinsic spectral cutoff weakens these limits by factors of 1.4--2.2, the weakening being smallest at the highest thresholds; with the factor computed at each threshold, the NGC~1068 constraint then remains the stronger one only for $m_{Z'}\gtrsim0.7$~GeV, by about 10\%.
For comparison, the event-count criterion of Sec.~\ref{sec:eventcount}, which fixes the intrinsic spectrum, gives limits on $g_\nu$ that are stronger by factors of 2--4. With this criterion, the broad-kernel constraints would be stronger than the laboratory bounds for $m_{Z'}\gtrsim0.23$~GeV, whereas the narrow-kernel constraints would remain weaker.
For the stellar-heated spike BM1$'$, the constraints are approximately an order of magnitude weaker than those for BM1 and remain less stringent than existing laboratory bounds for both kernel benchmarks.

We emphasize that $g_\nu$ is treated as an effective neutrino coupling in our analysis. As discussed above, a neutrino-specific vector interaction requires a gauge-invariant ultraviolet completion, which may introduce additional interactions and experimental constraints. Consequently, the relatively large couplings $g_\nu\sim10^{-2}$ relevant to the astrophysical limits must be assessed within a specific completion. Such a construction is beyond the scope of the present work, and our coupling constraints should be interpreted within the effective interaction framework.

Finally, we comment on the cosmology of $\chi_1$. For the mass hierarchy considered here, $m_1<m_{Z'},m_{h_D}$, the tree-level annihilation channels $\chi_1\chi_1\to Z'Z'$ and $\chi_1\chi_1\to h_Dh_D$ are kinematically forbidden, so the relic abundance cannot be set by standard freeze-out. At the same time, the couplings probed here would bring $\chi_1$ into equilibrium with the neutrino bath through $\nu\bar\nu\to\chi_1\chi_2$: for $m_1=300$~keV, $m_2=1$~GeV, $g_\nu\simeq0.02$, and $g_\chi\simeq0.05$, an order-of-magnitude estimate gives a rate above the Hubble rate down to $T\simeq m_2/20\approx50$~MeV, and a $\chi_1$ population thermalized in this way would overclose the Universe. A viable cosmology therefore requires a reheating temperature below $\sim m_2/20$, so that $\chi_1$ never thermalizes, yet high enough for Big Bang nucleosynthesis; the observed abundance must then be produced out of equilibrium, for example by freeze-in near the reheating temperature. We do not address this production history here; our transport calculation takes the $\chi_1$ density in the spike as given by the benchmark profiles.

\section{Conclusions}
\label{sec:conclusions}

Inelastic neutrino--dark-matter (DM) scattering introduces a nonzero minimum energy loss in neutrino propagation through DM spikes surrounding supermassive black holes. For a stationary ground-state DM target and an on-shell excited state, each scattering removes at least $\Eth=(m_2^2-m_1^2)/(2m_1)$ from the incident neutrino. Unlike elastic scattering mediated by a light particle, this finite energy loss prevents the near-cancellation between attenuation and regeneration for falling intrinsic neutrino spectra. By solving the full energy-space cascade, we have shown that both narrow and broad energy-transfer kernels produce a characteristic spectral deficit above threshold, accompanied by a pile-up below it. A broader kernel shifts the deficit toward higher energies and delays the high-energy recovery of the transmitted flux. Secondary neutrinos from prompt excited-state decays enhance the subthreshold pile-up, while the DM velocity distributions considered here broaden and slightly shift the spectral feature without erasing it.

Using the publicly available IceTracks-DR2 data for NGC~1068, we derive constraints on the asymptotic optical depth $\tau_0$ through an unbinned likelihood analysis, profiling over the intrinsic power-law normalization and spectral index. For the narrow kernel ($\epsilon=0.01$), we obtain $\tau_0<26$--39 over $\Eth=0.24$--3.0~TeV under the nominal 90\% confidence-level prescription $\Delta{\rm TS}=2.71$, with no upper limit for $\Eth\gtrsim3.5$~TeV. For the broader kernel ($\epsilon=0.36$), the constraints strengthen to $\tau_0<20$--26 over $\Eth=0.24$--2.4~TeV but weaken near 3~TeV, where the broadened spectral edge overlaps the energy range most relevant to the likelihood. 
These likelihood constraints are substantially weaker than those obtained from the illustrative fixed-flux event-count criterion, since variations in the intrinsic source spectrum can partially compensate for attenuation. Allowing an exponential cutoff further weakens the narrow-kernel limits by factors of 2.2--4.8 over $\Eth=0.3$--3.0~TeV (7.4 at $\Eth=0.12$~TeV) and the broad-kernel limits by factors of 2.0--5.0 over $\Eth=0.24$--2.4~TeV, highlighting their sensitivity to assumptions about the intrinsic spectrum.

We further extend the analysis to CGCG~420-015 and TXS~0506+056. Combining these sources with NGC~1068, with their optical depths related through the assumed DM spike columns, extends the threshold reach to approximately 12~TeV. This improvement is conditional on the additional neutrino excesses being astrophysical in origin and on the adopted relative column densities. The likelihood scans show no significant preference for a threshold-induced spectral feature over most of the parameter range. A modest improvement appears near $\Eth\simeq0.12$~TeV when the source optical depths are fitted independently, but it is substantially reduced when the depths are tied to a common particle-physics model.

Translating the optical-depth constraints into bounds on the scattering cross section and underlying couplings requires additional astrophysical and particle-physics assumptions. In particular, the DM column densities of the adiabatic (BM1) and stellar-heated (BM1$'$) spike benchmarks differ by a factor of approximately 650. In a vector-mediator realization, the kernel width is determined by $\epsilon=m_{Z'}^2/(m_2^2-m_1^2)$, while the associated dark-sector interactions can induce DM self-scattering and modify the inner spike profile. Imposing the requirement that the spike remain collisionless, we obtain $g_\nu\lesssim0.018\text{--}0.049$ for $m_1=300$~keV under the adiabatic spike assumption, with only a weak dependence on $\epsilon$ at fixed threshold. The kernel width nevertheless determines the mediator-mass range over which these constraints apply. For the narrow kernel ($m_{Z'}\simeq27$--135~MeV), the resulting limits are more than an order of magnitude weaker than existing bounds from $K$ decays. In contrast, for the broader kernel ($\epsilon=0.36$), the NGC~1068 constraints can be stronger than the laboratory bounds from $m_{Z'}\simeq0.45$~GeV up to the end of the scanned range (0.81~GeV), by up to a factor of 1.5; once an intrinsic spectral cutoff is allowed, this holds only marginally, above $m_{Z'}\simeq0.7$~GeV.
This sensitivity requires a dense adiabatic spike and an effective neutrino coupling whose ultraviolet realization remains to be established. For the stellar-heated spike, the constraints remain weaker than the laboratory bounds.

Our results demonstrate that the nonzero minimum energy loss is essential for interpreting high-energy neutrino observations as probes of inelastic dark sectors. The detailed shape of the energy-redistribution kernel matters at the tens-of-percent level in the derived limits for broad kernels and at the few-percent level for narrow ones, although the transmission just above threshold is more sensitive to it.
While the conversion to particle-physics couplings depends strongly on the assumed DM distribution and model realization, the threshold-induced spectral distortion provides a distinctive signature of inelastic neutrino--DM scattering.

\begin{acknowledgments}
 This work was supported in part by the National Science and Technology Council (NSTC) Grant No. NSTC-114-2112-M-002-020-MY3 (CWC), the Ministry of Education (Higher Education Sprout Project NTU-114L104022-1), the National Center for Theoretical Sciences of Taiwan (VQT), and the Vietnam National Foundation for Science and Technology Development (NAFOSTED) under grant number 103.01-2023.50 (VQT). 
This work made use of the public IceCube data release IceTracks-DR2 and the SkyLLH code. 
We also acknowledge the use of AI-assisted tools for coding assistance, grammar and spelling checks, and language polishing of the manuscript.
\end{acknowledgments}

\onecolumngrid

\appendix
\section{Particle-physics model}\label{app:model}
\subsection{A $U(1)_D$ realization of the inelastic dark sector}
\label{a:sec:dark_model}

We briefly describe a simple $U(1)_D$ realization of the phenomenological
interaction used in the main text. Consider two Weyl fermions $\eta$ and
$\xi$ with $U(1)_D$ charges $+1$ and $-1$, respectively, a complex
scalar $S$ with charge $+2$, and the $U(1)_D$ gauge boson $Z'$ with gauge coupling $g_\chi$. The dark-sector Lagrangian is
\begin{equation}
\mathcal L_D=-\frac14 Z'_{\mu\nu}Z'^{\mu\nu}+(D_\mu S)^\dagger D^\mu S-V(S) 
+i\eta^\dagger\bar\sigma^\mu D_\mu\eta+i\xi^\dagger\bar\sigma^\mu D_\mu\xi-\mathcal L_{\rm mass},
\label{a:eq:LD}
\end{equation}
where
\begin{equation}
D_\mu=\partial_\mu - i g_\chi Q_D Z'_\mu,
\qquad
Z'_{\mu\nu}=\partial_\mu Z'_\nu-\partial_\nu Z'_\mu,
\end{equation}
$Q_D=(+1,-1,+2)$ for $(\eta,\xi,S)$, the scalar potential $V(S)= -\mu_S^2|S|^2+\lambda|S|^4$, and the mass terms
\begin{equation}
-\mathcal{L}_{\rm mass}
=
m_D\eta\xi
+\frac{y_S}{2}S^\dagger\eta\eta
+\frac{y_S}{2}S\xi\xi
+\mathrm{h.c.},
\end{equation}
where an exchange symmetry $\eta\leftrightarrow\xi$,
$S\leftrightarrow S^\dagger$ has been imposed, and $\bar\sigma^\mu=(1,-\vec\sigma)$ with $\vec\sigma$ the Pauli matrices. The fermion kinetic terms contain the gauge interaction
$g_\chi Z'_\mu(\eta^\dagger\bar\sigma^\mu\eta-\xi^\dagger\bar\sigma^\mu\xi)$. After $S$ develops
a vacuum expectation value $v_D$, $S = (v_D + h_D) /\sqrt{2}$ in unitary gauge, the gauge boson and the dark Higgs ($h_D$) acquire masses
\begin{equation}
m_{Z'}=2g_\chi v_D,\qquad m_{h_D}=\sqrt{2\lambda}\,v_D,
\end{equation}
and the fermion mass matrix is
\begin{equation}
\mathcal{M}=
\begin{pmatrix}
m_M&m_D\\
m_D&m_M
\end{pmatrix},
\qquad
m_M=\frac{y_S v_D}{\sqrt{2}}.
\end{equation}
Its eigenvalues are $m_D+m_M$ for $(\eta+\xi)/\sqrt2$ and $m_M-m_D$ for $(\eta-\xi)/\sqrt2$. We take $m_D>m_M$, so the second eigenvalue is
negative and is made positive by a phase rotation. The two-component mass eigenstates are
\begin{equation}
\psi_1=\frac{i}{\sqrt{2}}(\xi-\eta),
\qquad
\psi_2=\frac{1}{\sqrt{2}}(\eta+\xi),
\end{equation}
with masses
\begin{equation}
m_1=m_D-m_M,
\qquad
m_2=m_D+m_M,
\qquad m_2-m_1=2m_M,
\label{a:eq:hierarchical_masses}
\end{equation}
and we denote by $\chi_{a}=(\psi_a,\psi_a^\dagger)^T$ ($a=1,2$) the corresponding four-component Majorana spinors. Consequently, the $U(1)_D$ current is purely off-diagonal, $\mathcal{L} \supset i g_\chi Z'_\mu\bar{\chi}_2\gamma^\mu\chi_1$,
where the overall sign depends on the phase convention of $\psi_1$. For $m_M>m_D$ no phase rotation is needed and the current is instead
axial vector. We focus on the vector interaction for the analysis. 
The dark Higgs couples diagonally to the Majorana fermions and to $Z'$,
\begin{equation}
\label{a:eq:darkhiggs}
\mathcal{L}\supset
-\frac{y_S}{2\sqrt{2}}h_D(\bar{\chi}_2\chi_2-\bar{\chi}_1\chi_1)
+\frac{m_{Z'}^2}{v_D}h_D Z'_\mu Z'^{\mu},
\end{equation}
and its Yukawa coupling is fixed by the masses and the gauge coupling,
\begin{equation}
\label{a:eq:gxyS}
\frac{y_S}{\sqrt{2}} = \frac{m_M}{v_D} = g_\chi\frac{m_2-m_1}{m_{Z'}}.
\end{equation}
For $m_1=300$~keV and $m_2=1$~GeV, Eq.~\eqref{a:eq:hierarchical_masses} requires $m_D=500.15$~MeV and $m_M=499.85$~MeV. The spectrum is therefore
not in the conventional pseudo-Dirac limit $m_D\gg m_M$; it requires $m_D\simeq m_M$, whose origin we
do not address. 

Finally, at the phenomenological level we parametrize the coupling of $Z'$ to neutrinos by the effective interaction
\begin{equation}
\mathcal{L}_{\nu Z'}\supset g_\nu Z'_\mu\bar{\nu}\gamma^\mu P_L\nu .
\end{equation}
The relevant interactions are therefore
\begin{equation}
\mathcal{L}\supset i g_\chi Z'_\mu\bar{\chi}_2\gamma^\mu\chi_1+g_\nu Z'_\mu\bar{\nu}\gamma^\mu P_L\nu,
\label{a:eq:L}
\end{equation}
as used in the main text. 
We note that the Majorana bilinears obey $\bar\chi_a\gamma^\mu\chi_b = -\bar\chi_b\gamma^\mu\chi_a$. Hence the first term in Eq.~\eqref{a:eq:L} is Hermitian by itself.

\subsection{Scattering cross section}\label{a:sec:xs}
We use the Feynman rules of Ref.~\cite{Denner:1992vza}:
for the fermion flow $\chi_1\to\chi_2$ the vertex is $i(ig_\chi\gamma^\mu)$ between $\bar u(p')$ and $u(p)$; for the reversed flow it is $\mathcal C\gamma^{\mu T}\mathcal C^{-1}=-\gamma^\mu$, with $\mathcal C$ the charge-conjugation matrix, 
between $\bar v(p)$ and $v(p')$. The process $\nu\chi_1\to\nu\chi_2$ proceeds through a single $t$-channel
diagram with no symmetry factor, and both fermion-flow choices give the same $|\mathcal M|^2$. The $q_\mu q_\nu/m_{Z'}^2$ term of the propagator, with $q=k-k'$,
drops because the massless-neutrino current is conserved, and the antisymmetric part of the lepton tensor contracts with a symmetric
DM tensor, so $\nu$ and $\bar\nu$ give identical results. Averaging over the initial $\chi_1$ spin, the squared matrix element is given in Eq.~{\eqref{eq:msq}} of the main text. 
In the forward-scattering limit, $t\to0$, we have $\overline{|\mathcal M|^2}=4g_\nu^2g_\chi^2(s-m_1^2)(s-m_2^2)/m_{Z'}^4>0$.

In the lab frame with $\chi_1$ at rest, $Q^2 = -t = 2m_1(E_\nu - E'_\nu-\Eth)$, with $E'_\nu$ the outgoing neutrino energy, and the differential cross section is
\begin{align}
\frac{d\sigma}{dQ^2} &= \frac{1}{16 \pi \left( s - m_1^2 \right)^2} \overline{|\mathcal M|^2} \nonumber\\
&= \frac{g_\nu^2g_\chi^2}{8\pi E_\nu^2(Q^2+m_{Z'}^2)^2}\Big[E_\nu(E_\nu-\Eth) +E'_\nu (E'_\nu+\Eth)-\frac{m_2Q^2}{2m_1}\Big],
\label{a:eq:dsigmadQ2}
\end{align}
where $0\le Q^2\le Q^2_{\max}= 4m_1E_\nu (E_\nu -\Eth)/(m_1+ 2 E_\nu)$.

Integrating Eq.~\eqref{a:eq:dsigmadQ2} over the momentum transfer, one obtains the cross section
\begin{equation}
\sigma(E_\nu)=\frac{g_\nu^2g_\chi^2}{16\pi m_1E_\nu^2}\left[\delta\omega_m-(2\mu+b)\ln\!\Big(1+\frac{\delta\omega_m}{\mu}\Big)+\frac{c\,\delta\omega_m}{\mu(\delta\omega_m+\mu)}\right],
\label{a:eq:sigma}
\end{equation}
where $\delta \omega_m=Q^2_{\max}/2m_1$ is the maximal excess loss, 
$\mu=m_{Z'}^2/2m_1=\epsilon\Eth$, $b=2E_\nu-\Eth+m_2$ and
$c=2E_\nu(E_\nu-\Eth)+b\mu+\mu^2$.
For $(E_\nu-\Eth)/\Eth\gg\epsilon\equiv m_{Z'}^2/(m_2^2-m_1^2)$ (in practice, $(E_\nu-\Eth)/\Eth\gtrsim80\,\epsilon$ for 5\% accuracy), Eq.~\eqref{a:eq:sigma} approaches
\begin{equation}
\label{a:eq:sigmaA}
\sigma(E_\nu) \simeq \frac{g_\nu^2g_\chi^2}{4\pi m_{Z'}^2}\Big(1-\frac{\Eth}{E_\nu}\Big).
\end{equation}

\begin{figure}[htbp]\centering\includegraphics[width=0.95\textwidth]{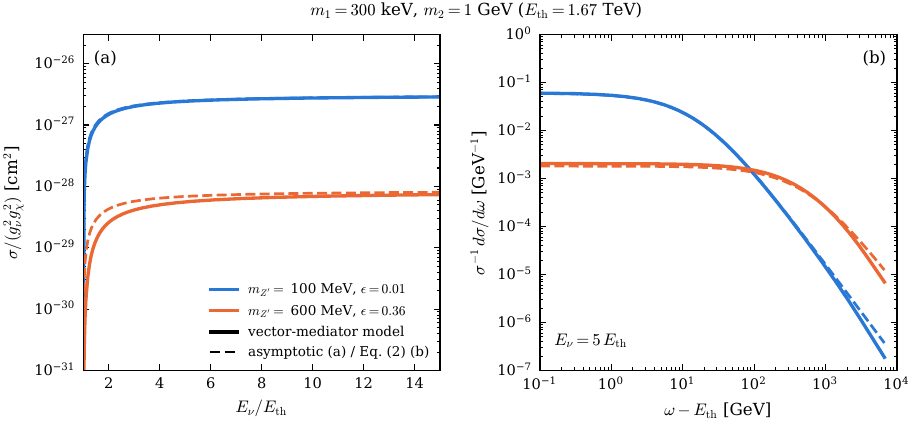}
\caption{
(a) Coupling-normalized cross section as a function of $E_\nu/\Eth$. The solid curves show the full vector-mediator result from Eq.~\eqref{a:eq:sigma}, while the dashed curves represent the asymptotic approximation in Eq.~\eqref{a:eq:sigmaA}.
(b) Normalized energy-transfer kernel $K=\sigma^{-1}d\sigma/d\omega$ at $E_\nu=5\Eth$ as a function of the excess energy loss $\omega-\Eth$. The solid curves show the full vector-mediator result from Eq.~\eqref{a:eq:dsigmadE}, while the dashed curves correspond to the approximate kernel in Eq.~\eqref{eq:sigma}.
In both panels, we fix $m_1=300$~keV and $m_2=1$~GeV ($\Eth=1.67$~TeV), and consider $m_{Z'}=100$ and 600~MeV, corresponding to $\epsilon=0.01$ and 0.36, respectively.
}
\label{a:fig:xs}
\end{figure}

Figure~\ref{a:fig:xs}(a) compares the full and asymptotic cross sections as functions of $E_\nu/\Eth$. As expected, deviations from the asymptotic approximation become more pronounced at lower neutrino energies and for larger mediator masses $m_{Z'}$, corresponding to broader energy-transfer kernels (larger $\epsilon$). For $\epsilon=0.36$, the full cross section lies 11\% below Eq.~\eqref{a:eq:sigmaA} at $E_\nu=10\Eth$ and 5\% below at $30\Eth$; for $\epsilon=0.01$, the deviation is below 1\% for $E_\nu\ge10\Eth$.

At fixed $E_\nu$, $Q^2$ is linear in the outgoing energy, so the energy-transfer kernel of the cascade equation,
$K(E_\nu\!\to\!E'_\nu)=\sigma(E_\nu)^{-1}d\sigma(E_\nu \to E'_\nu)/dE'_\nu$, follows with the Jacobian $|dQ^2/dE'_\nu|=2m_1$:
\begin{equation}
\frac{d\sigma(E_\nu\to E'_\nu)}{dE'_\nu}=2m_1\,\frac{d\sigma}{dQ^2}\bigg|_{Q^2=2m_1(E_\nu-E'_\nu-\Eth)},
\label{a:eq:dsigmadE}
\end{equation}
where $E_\nu-\Eth-{Q^2_{\max}}/{(2m_1)}\le E'_\nu\le E_\nu-\Eth$,
and it vanishes outside this range, in particular for losses below $\Eth$.

Figure~\ref{a:fig:xs}(b) compares the energy-transfer kernel $K(E_\nu\to E_\nu')$ at $E_\nu=5\Eth$ with the phenomenological approximation of Eq.~\eqref{eq:sigma}, as functions of the excess energy loss $\omega-\Eth$. The kernel is maximal at threshold and falls to about one quarter of its threshold value at $\omega-\Eth=m_{Z'}^2/(2m_1)$, where $Q^2=m_{Z'}^2$, and as $(\omega-\Eth)^{-2}$ for $\omega-\Eth\gg m_{Z'}^2/(2m_1)$, with only a weak dependence on the incident neutrino energy. Consequently, for a narrow kernel with $\epsilon=0.01$, the energy loss is concentrated near its kinematic minimum, yielding $\langle\omega\rangle=(1.03$--$1.07)\Eth$ over the range $E_\nu=3$--100~TeV.

All results in the main text use the analytic expressions of Eqs.~\eqref{a:eq:dsigmadQ2}--\eqref{a:eq:sigma}, while the asymptotic form
\eqref{a:eq:sigmaA} only defines $\sigma_0$ and $\tau_0$.

\subsection{The decay $\chi_2\to\chi_1\nu\bar\nu$}

The excited DM state $\chi_2$ decays through the on-shell cascade $\chi_2\to\chi_1Z'$, followed by $Z'\to\nu\bar\nu$. The corresponding decay widths are
\begin{equation}
\begin{split}
\Gamma(\chi_2\to\chi_1Z')
={}&\frac{p^*g_\chi^2}{16\pi m_2^2}
\Bigg[
8p_1\cdot p_2-16m_1m_2
+\frac{4}{m_{Z'}^2}
\left[
2(k\cdot p_1)(k\cdot p_2)
-m_{Z'}^2(p_1\cdot p_2-m_1m_2)
\right]
\Bigg],\\
\Gamma(Z'\to\nu\bar\nu)
={}&\frac{3g_\nu^2m_{Z'}}{24\pi},
\end{split}
\label{a:eq:width}
\end{equation}
where $p_2$ and $p_1$ are the four-momenta of the initial $\chi_2$ and final $\chi_1$, respectively, and $k=p_2-p_1$ is the four-momentum of the emitted $Z'$. The quantity $p^*$ denotes the magnitude of the three-momentum of either daughter particle in the $\chi_2$ rest frame. 
We consider couplings for which both decays occur promptly on the relevant propagation scales. 
We also treat $\chi_2$ as a narrow state (i.e., {$\Gamma_{\chi_2}\ll m_2$}).
Since the excited states are highly boosted in the scattering process, the resulting secondary neutrinos are approximately collinear with the incident neutrino, with opening angles of order $m_2/E_2\le m_2/\Eth\simeq0.03^\circ$ for the benchmark $m_2=1$~GeV, well below the IceCube angular resolution.

To describe the secondary-neutrino spectrum, we adopt the approximation of isotropic $Z'$ decays. In the vector-mediator model, however, the emitted $Z'$ is predominantly longitudinally polarized, with a longitudinal fraction $F_L\simeq m_2^2/(m_2^2+2m_{Z'}^2)$, giving $F_L\simeq0.98$ for $m_{Z'}=100$~MeV and $m_2=1$~GeV, and $F_L\simeq0.58$ for $m_{Z'}=600$~MeV. Accounting for spin correlations yields a narrower $\nu+\bar\nu$ spectrum with a suppressed hard tail, proportional to $6(1-x)^2$ rather than $-2\ln x$ in the limit $m_1\ll m_{Z'}\ll m_2$; both spectra have the same mean, $\langle x\rangle=1/4$ per neutrino. Here $x=E_\nu/E_2$ is the fractional energy carried by a secondary neutrino, with $E_2=m_1+\omega$ denoting the energy of the excited DM state in the laboratory frame. Replacing the isotropic approximation with the spin-correlated spectrum lowers the pile-up below threshold for $\epsilon=0.01$, from $R(0.9\Eth)=1.24$ to 1.17 at $\tau_0=30$, but changes the transmission above $1.5\Eth$ by at most 2.4\% for $\tau_0\le100$ and the NGC~1068 likelihood limits by factors of 0.98--1.01 at $\Eth=0.12$--2.4~TeV. For $\epsilon=0.36$, where the $Z'$ is less polarized, the transmission changes by at most 1.1\% for $\tau_0\le100$, both below and above threshold, and the number of signal neutrinos in the 1.5--15~TeV window by at most 1.3\%.

Within the isotropic approximation, the energy distribution of secondary neutrinos from each $\chi_2$ decay is
\begin{equation}
f_{\nu\bar\nu}(x)=
\frac{2}{x_+-x_-}
\ln\frac{x_+}{\max(x,x_-)},
\qquad 0\le x<x_+,
\label{a:eq:fsec}
\end{equation}
and vanishes otherwise. The kinematic endpoints are $x_\pm=(E_{Z'}^*\pm p^*)/m_2$, where
\begin{equation}
E_{Z'}^*=\frac{m_2^2+m_{Z'}^2-m_1^2}{2m_2},
\qquad
p^*=\sqrt{(E_{Z'}^*)^2-m_{Z'}^2}.
\end{equation}
The distribution is normalized to two secondary neutrinos per decay, $\int dx\,f_{\nu\bar\nu}(x)=2$. For an unpolarized $\chi_2$, the neutrino pair carries an average fraction $E_{Z'}^*/m_2\simeq50.5\%$ of the excited-state energy for $m_{Z'}=100$~MeV (68\% for $m_{Z'}=600$~MeV). We have verified the analytical spectrum against a Monte Carlo simulation of the sequential decays with isotropic $Z'$ decay angles, finding agreement at the 5\% level.

Convolving the secondary spectrum with the scattering kernel gives
\begin{equation}
K_{\rm sec}(\widetilde E_\nu\to E_\nu)
=\int_{\Eth}^{\widetilde E_\nu}d\omega\,
K(\widetilde E_\nu\to\widetilde E_\nu-\omega)
\,\frac{1}{E_2}
f_{\nu\bar\nu}\left(\frac{E_\nu}{E_2}\right),
\label{a:eq:Ksec}
\end{equation}
where $E_2=m_1+\omega$. The kernel is normalized such that $\int dE_\nu\,K_{\rm sec}=2$, corresponding to one neutrino and one antineutrino produced per scattering. In the kinematic regime considered here, $E_\nu\le x_+E_2<\widetilde E_\nu$, so the secondary neutrinos remain less energetic than the incident neutrino, justifying the energy ordering adopted in the transport equation.

Finally, the on-shell decay $\chi_2\to\chi_1Z'$ requires $m_{Z'}<m_2-m_1$. For heavier mediators, $\chi_2$ can instead undergo the three-body decay $\chi_2\to\chi_1\nu\bar\nu$ through an off-shell $Z'$. Since the resulting decay spectrum and lifetime differ from those of the on-shell cascade, we do not consider this regime.

\section{Neutrino transport}\label{app:transport}
\label{a:sec:transport}

We compute the transmission factor $R=\Phi^{\rm out}/\Phi^{\rm in}$ by numerically solving the cascade equation, Eq.~\eqref{eq:dflux}. We discretize the neutrino energy into 360 logarithmically spaced bins with edges $e_j$, covering $10^{-3}\le E_\nu/\Eth\le10^4$ ($\Delta\ln E\simeq0.045$). The state vector contains the bin-integrated fluxes,
$\Phi_j(X)=\int_{e_j}^{e_{j+1}}\Phi(E_\nu;X)\,dE_\nu$,
with initial conditions $\Phi_j^{\rm in}=\Phi_j(0)$ obtained from the intrinsic power-law spectrum $\Phi^{\rm in}\propto E_\nu^{-\gamma}$.

Within each incoming energy bin $i$, we evaluate the incident-energy dependence using quadrature nodes $\widetilde E_{i,n}$ and normalized weights $w_{i,n}$, constructed using trapezoidal integration in $\widetilde E-e_i$ with an in-bin spectral weight proportional to $\widetilde E^{-3}$. The nodes are distributed logarithmically near the lower bin edge, where transitions into lower-energy bins originate, and near $\widetilde E=e_i+\Eth$, where the energy-loss probability varies rapidly over a characteristic scale $m_{Z'}^2/(2m_1)$, as illustrated in Fig.~\ref{a:fig:xs}(b). We supplement these with uniformly spaced nodes, using 64 nodes for each family. This adaptive sampling suppresses numerical aliasing when the energy-bin width exceeds the inelastic threshold.

The loss coefficients {$\hat\sigma_i$}, primary-regeneration matrix {$\hat K_{ji}$}, and secondary-neutrino matrix {$\hat K^{\rm sec}_{ji}$} are constructed as
\begin{align}
{\hat\sigma_i}={}&\sum_n w_{i,n}\,\sigma(\widetilde E_{i,n}),
\nonumber\\
{\hat K_{ji}}={}&\sum_n w_{i,n}
\left[
\sigma\big(\widetilde E_{i,n};<Q^2_j\big)
-\sigma\big(\widetilde E_{i,n};<Q^2_{j+1}\big)
\right],
\label{a:eq:Tji}\\
{\hat K^{\rm sec}_{ji}}={}&\sum_n w_{i,n}\sum_q
\Delta\sigma_q(\widetilde E_{i,n})
\left[
F_{\nu\bar\nu}\left(\frac{e_{j+1}}{E_{2,q}}\right)
-F_{\nu\bar\nu}\left(\frac{e_j}{E_{2,q}}\right)
\right],
\nonumber
\end{align}
where $Q_j^2=2m_1(\widetilde E_{i,n}-e_j-\Eth)$, and $\sigma(E;<Q_c^2)$ 
denotes the cross section integrated over momentum transfers $Q^2<Q_c^2$, 
obtained analytically from Eq.~\eqref{a:eq:sigma} 
by replacing $\delta\omega_m$ with $Q_c^2/(2m_1)$. 
For the secondary contribution, we divide the momentum transfer into 
48 logarithmic $Q^2$ intervals labeled by $q$. 
Here $\Delta\sigma_q$ is the cross section integrated over interval $q$, $E_{2,q}=m_1+\Eth+\bar Q_q^2/(2m_1)$ is the corresponding excited-state energy, with $\bar Q_q^2$ the midpoint of the interval, 
and $F_{\nu\bar\nu}(x)=\int_0^x f_{\nu\bar\nu}(x')\,dx'$ is the cumulative secondary-neutrino spectrum. The cumulative cross section is set to zero for $Q_c^2\le0$ and to $\sigma(\widetilde E_{i,n})$ for $Q_c^2\ge Q^2_{\max}$, so that outgoing bins outside the kinematic range receive no flux. Each scattering moves the incident neutrino to a lower energy, possibly within the same bin, and adds two secondaries, so that $\sum_j\hat K_{ji}=\hat\sigma_i$ and $\sum_j\hat K^{\rm sec}_{ji}=2\hat\sigma_i$, up to the part of the outgoing spectrum that falls below the lower grid edge. Numerically, $\sum_j\hat K_{ji}/\hat\sigma_i\ge0.998$ (0.990) and $\sum_j\hat K^{\rm sec}_{ji}/\hat\sigma_i=1.95$--2.00 for $E_\nu\ge1.05\,\Eth$ at $\epsilon=0.01$ (0.36).

Analytically integrating the cross section over each outgoing energy bin preserves the transition probability even when the energy-transfer kernel is narrower than the bin width, although spectral structure within individual bins remains unresolved. To assess numerical convergence, we double the number of energy bins and compare the transmission averaged over 0.1-dex energy intervals, for $m_1=300$~keV, $\Eth=1.67$~TeV, $\gamma=3.2$, and $0.05\le E_\nu/\Eth\le10^3$. The resulting changes are at most 0.4\% for $\tau_0\le10$ and 4\% for $\tau_0\le100$ wherever $R>10^{-2}$ for $\epsilon=0.01$, and at most 0.13\% and 3\% for $\epsilon=0.36$. The largest deviation, approximately 40\% at $\tau_0=100$ (9\% for $\epsilon=0.36$), occurs in the interval containing the spectral edge, where the transmission is strongly suppressed ($R<10^{-2}$).

Defining the transport matrix $A=-{\rm diag}(\hat\sigma)+\hat K+\hat K^{\rm sec}$, the discretized cascade equation becomes $d\vec\Phi/dX=A\vec\Phi$, with the solution
\begin{equation}
\vec\Phi(X_{\rm tot})=e^{AX_{\rm tot}}\vec\Phi^{\rm in},
\qquad
R_j=\frac{\Phi_j(X_{\rm tot})}{\Phi_j^{\rm in}},
\qquad
X_{\rm tot}=\frac{\tau_0}{\sigma_0}.
\label{a:eq:expm}
\end{equation}
The matrix exponential provides the exact solution of the discretized transport equation. 
For the likelihood analysis, we construct transmission tables at 31 values of $\tau_0$ spanning $10^{-2}$--$10^3$, with successive values separated by a factor of approximately 1.47, and 23 values of the intrinsic spectral index $\gamma$ spanning 2--4 (extended to $\gamma\ge1.5$ for TXS~0506+056; Sec.~\ref{sec:other}). The likelihood is evaluated at the tabulated optical depths, and only the crossing defining the nominal 90\% CL upper limit is interpolated in $\ln\tau_0$. 
Replacing linear interpolation with monotone cubic interpolation changes the resulting limits by at most 1\%. Transmission curves at specific optical depths, such as those in Fig.~\ref{fig:f}(c), are computed directly from Eq.~\eqref{a:eq:expm}.

We validate the transport solver against an independent event-by-event Monte Carlo simulation using $2\times10^6$ neutrinos per benchmark point. The simulation employs the same analytical total cross section $\sigma(E_\nu)$ as the cascade solver, while sampling the momentum transfer $Q^2$ from its cumulative distribution. This comparison therefore tests the numerical treatment of neutrino propagation independently of the matrix-exponentiation method, while the underlying scattering matrix element is validated separately in Sec.~\ref{a:sec:xs}. For $\Eth=1.67$~TeV, $\epsilon=0.01$, and $\gamma=3.2$, considering primary neutrinos only and using the production energy grid, we find agreement at $\tau_0=1$, 3, 10, and 30, with $\chi^2/{\rm ndf}\le29/32$ and relative deviations below 2\% outside statistically limited bins.

\begin{figure}[htbp]\centering\includegraphics[width=0.55\textwidth]{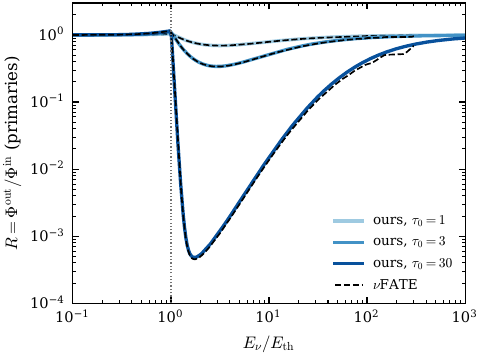}
\caption{Validation against $\nu$FATE~\cite{Vincent:2017svp} ($\Eth=1.67$~TeV, $\epsilon=0.01$, primaries only, $\Phi^{\rm in}\propto E_\nu^{-3.2}$, $0.05\le E_\nu/\Eth\le3000$).
Transmission from our solver (solid) and from $\nu$FATE with 3200 nodes and the cell-integrated kernel (dashed, shown up to
$300\,\Eth$), at $\tau_0=1$, 3, and 30.}\label{a:fig:nufate}
\end{figure}

As an additional cross-check, we solve Eq.~\eqref{eq:dflux} for primary neutrinos only using the public code $\nu$FATE~\cite{Vincent:2017svp}, supplying our total cross section and energy-transfer kernel on its logarithmic energy grid over $0.05\le E_\nu/\Eth\le3000$. With 3200 energy nodes in $\nu$FATE, the two implementations agree within 1\% for $\tau_0=1$ and 3\% for $\tau_0=3$ over $1.2\le E_\nu/\Eth\le300$. At $\tau_0=30$, the agreement remains within 9\% over $E_\nu/\Eth=1.5$--10. The comparison is summarized in Fig.~\ref{a:fig:nufate}.

\section{Multi-source limits on the optical depth}
\label{app:multi}

For the multi-source analysis of Sec.~\ref{sec:other}, we extend the DM spike prescriptions and host-halo relations adopted for NGC~1068 to CGCG~420-015 and TXS~0506+056. For CGCG~420-015, we assume coronal neutrino emission at $R_{\rm em}=30R_S$ and a black-hole influence radius $r_h=6.5\times10^5R_S$, comparable to that of NGC~1068 ($r_h=0.65$~pc $=6.8\times10^5R_S$). For TXS~0506+056, we instead adopt jet emission at $R_{\rm em}=2\times10^{17}$~cm~\cite{Cline:2022qld}, corresponding to approximately $2.2\times10^3R_S$, while keeping the same ratio $r_h/R_S$. The adopted source parameters, DM column densities, and their ratios $\kappa=\Sigma/\Sigma_{\rm NGC\,1068}$, which determine the relative optical depths in Eq.~\eqref{eq:TSjoint}, are summarized in Table~\ref{a:tab:sources}.

For self-similar spike profiles with identical $R_{\rm em}/R_S$ and $r_h/R_S$, the column density scales as $\Sigma\propto M_{\rm BH}^{-1}$, independently of the spike slope $\alpha$. Consequently, the column-density ratio for CGCG~420-015 is approximately $\kappa\simeq M_{\rm NGC\,1068}/M_{\rm CGCG}=0.05$ for both BM1 and BM1$'$. A larger emission radius further suppresses the column density, approximately as $\Sigma\propto(R_{\rm em}/R_S)^{1-\alpha}$ for a power-law profile; this scaling neglects the factor $(1-4R_S/r)^3$, overestimates the suppression from $30R_S$ to $100R_S$ by 18\% (BM1) and 9\% (BM1$'$), and fails below ${\sim}30R_S$. The values of $\kappa$ in Table~\ref{a:tab:sources} are exact integrals. This explains the substantially smaller column-density ratio $\kappa$ of TXS~0506+056 and the stronger dependence of $\kappa$ on the assumed spike slope: $\kappa=1.4\times10^{-4}$ for BM1 and $4.5\times10^{-3}$ for BM1$'$, whereas $\kappa$ for CGCG~420-015 changes by only about 4\%.

\begin{table}[htbp]
\centering
\begin{tabular}{lccc}
\hline\hline
 & NGC~1068 & CGCG~420-015 & TXS~0506+056\\
\hline
$z$ & 0.0034 & 0.0296 & 0.3365~\cite{Paiano:2018qeq}\\
$M_{\rm BH}$ [$M_\odot$] & $1.0\times10^7$~\cite{Cline:2023tkp} & $2.0\times10^8$~\cite{Koss:2017ilk} & $3.09\times10^8$~\cite{Cline:2022qld}\\
$R_S$ [pc] & $9.6\times10^{-7}$ & $1.9\times10^{-5}$ & $3.0\times10^{-5}$\\
$r_h$ & 0.65~pc ($6.8\times10^5R_S$) & $6.5\times10^5R_S$ & $6.5\times10^5R_S$\\
$R_{\rm sp}$ [kpc] & 0.68 & 1.15 & 1.24\\
$R_{\rm em}$ & $30R_S$ & $30R_S$ & $2.2\times10^3R_S$\\
$\Sigma_{\rm BM1}$ [GeV\,cm$^{-2}$] & $5.6\times10^{31}$ & $2.9\times10^{30}$ & $7.7\times10^{27}$\\
$\Sigma_{\rm BM1'}$ [GeV\,cm$^{-2}$] & $8.7\times10^{28}$ & $4.7\times10^{27}$ & $3.9\times10^{26}$\\
$\kappa_{\rm BM1}$ & 1 & 0.0515 & $1.4\times10^{-4}$\\
$\kappa_{\rm BM1'}$ & 1 & 0.0535 & $4.5\times10^{-3}$\\
$\kappa_{\rm BM1}$, coronae at $100R_S$ & 1 & 0.0515 & $5.8\times10^{-4}$\\
\hline\hline
\end{tabular}
\caption{
Source parameters, DM column densities $\Sigma$, and column-density ratios $\kappa=\Sigma/\Sigma_{\rm NGC\,1068}$ adopted in the multi-source analysis. The columns are calculated for non-annihilating spike profiles, integrated from $R_{\rm em}$ to 100~pc. The redshifts of NGC~1068 and CGCG~420-015 are inferred from their distances of 14.4 and 127~Mpc~\cite{Cline:2023tkp,Inoue:2024nap}, assuming $H_0=70$~km\,s$^{-1}$\,Mpc$^{-1}$. The last row corresponds to $R_{\rm em}=100R_S$ for both coronal sources, while the emission radius of TXS~0506+056 remains unchanged.}
\label{a:tab:sources}
\end{table}

\begin{figure}[htbp]
\centering
\includegraphics[width=0.55\textwidth]{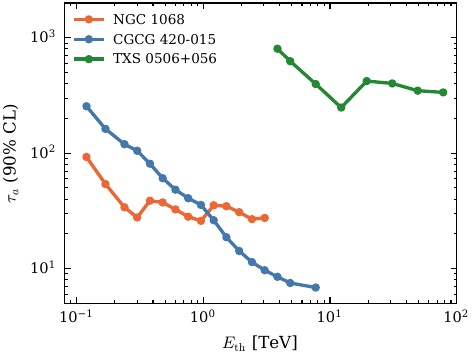}
\caption{
Nominal 90\% CL pointwise upper limits on the asymptotic optical depth $\tau_a$ for individual IceCube sources, normalized to their respective DM columns, as functions of the source-frame inelastic threshold $\Eth$.}
\label{a:fig:multi}
\end{figure}

Figure~\ref{a:fig:multi} presents the individual upper limits on the optical depth of each source, normalized to its own DM column density, obtained from the same SkyLLH likelihood scans used in Fig.~\ref{fig:dTS}. For CGCG~420-015, we obtain $\tau_a<255$ at $\Eth=0.12$~TeV and $\tau_a<105$ at 0.3~TeV. The constraint strengthens with increasing threshold, reaching $\tau_a<6.9$ at $\Eth=7.7$~TeV, while no upper limit is obtained above approximately 10~TeV. In contrast, TXS~0506+056 yields substantially weaker constraints, with $\tau_a<250$--800 over $\Eth=3.8$--78~TeV. Extending the optical-depth scan to $\tau_a=10^5$ gives upper limits of approximately $\tau_a=(3$--$4.5)\times10^3$ for $\Eth=0.3$--3~TeV. These individual constraints, together with the column-density ratios in Table~\ref{a:tab:sources}, determine the combined limits presented in Sec.~\ref{sec:other}.

\section{Bounds on $g_\chi$}
\label{app:selfconsistent}

The dark-sector coupling $g_\chi$ is constrained by perturbative unitarity and by DM self-interactions, which can modify the spike density profile assumed in our neutrino transport analysis. In the dark-Higgs completion of Appendix~\ref{app:model}, both constraints can be expressed in terms of the Yukawa coupling
\begin{equation}
y\equiv\frac{y_S}{\sqrt2}
=g_\chi\frac{m_2-m_1}{m_{Z'}}
\simeq\frac{g_\chi}{\sqrt\epsilon},
\label{a:eq:y}
\end{equation}
where the last relation holds for $m_1\ll m_2$. Thus, an upper bound on $y$ directly translates into a constraint on $g_\chi$.

\emph{Perturbative unitarity.}
We impose tree-level partial-wave unitarity on $\chi_a\chi_a\to Z'Z'$ ($a=1,2$), including $t$- and $u$-channel fermion exchange and $s$-channel dark-Higgs exchange. The partial-wave amplitudes are~\cite{Kahlhoefer:2015bea}
\begin{equation}
a_J^{fi}(s)=
\frac{\sqrt{\beta_i\beta_f}}{64\pi}
\int_{-1}^{1}d\cos\theta^*\,
d^J_{\zeta\,\zeta'}(\theta^*)\,
\mathcal M_{fi}(s,\theta^*),
\label{a:eq:aJ}
\end{equation}
where $J$ is the total angular momentum of the partial wave ($J=0,1,2$ are included), $\zeta=\lambda_1-\lambda_2$ and $\zeta'=\lambda_3-\lambda_4$ denote the total spin of the initial and final states along the collision axis~\cite{Kahlhoefer:2015bea} ($\lambda_i$ being the individual helicities), $\theta^*$ is the scattering angle in the center-of-mass frame, $\beta_{i,f}$ are the initial- and final-state velocity factors, $d^J_{\zeta\zeta'}$ are Wigner functions, and the normalization accounts for identical particles. Requiring the largest eigenvalue magnitude of the Hermitian part of the coupled-channel partial-wave matrix to satisfy $|\mathrm{Re}\,a_J|\le1/2$ gives an upper bound on $y$. Since the tree-level amplitudes scale as $y^2$, this can be written as
\begin{equation}
y\le y_{\max}
=\left[2\max_{s,J}\hat a_J(s)\right]^{-1/2},
\label{a:eq:ymax}
\end{equation}
where $\hat a_J$ denotes the relevant maximal partial-wave magnitude evaluated at $y=1$, away from resonance poles. For the mass hierarchies considered here,  the strongest constraint arises in the $J=0$ coupled channel $\chi_1\chi_1,\chi_2\chi_2\to Z'Z'$ at $\sqrt{s}\simeq2.9$~GeV $\approx3m_2$, above the $\chi_2\chi_2$ threshold $\sqrt s=2m_2$, where the $\chi_2\chi_2$ amplitude adds to the $\chi_1\chi_1$ one. At higher energies the $J=0$ amplitudes, which require a helicity flip, decrease with $\sqrt s$, while the $J=2$ amplitude only approaches $\sqrt3\,y^2/(48\pi)$, half the $J=0$ maximum, yielding the approximate bound
\begin{equation}
g_\chi\lesssim4.6\,
\frac{m_{Z'}}{m_2-m_1}
\simeq4.6\sqrt\epsilon.
\label{a:eq:gchimax}
\end{equation}
As shown below, the requirement of a collisionless DM spike imposes a substantially stronger constraint.

\emph{DM self-scattering.}
In the dark-Higgs completion, elastic $\chi_1\chi_1$ scattering is mediated by $h_D$ exchange. For $m_{h_D}\gg m_1v$, with $v$ the DM velocity, the interaction reduces to a scalar contact operator with coefficient $C=y^2/m_{h_D}^2$. In the Born approximation, the viscosity cross section for identical Majorana fermions is~\cite{Tulin:2017ara,Colquhoun:2020adl}
\begin{equation}
\sigma_V= \frac12\int d\Omega\,\sin^2\theta^*\,\frac{d\sigma_{\chi_1 \chi_1 \to \chi_1 \chi_1}}{ d\Omega} = \frac{C^2m_1^2}{12\pi},
\end{equation}
where the factor $1/2$ avoids double counting of identical final states.
Using $m_{h_D}^2=2\lambda v_D^2$, $v_D=(m_2-m_1)/(2y)$, and the tree-level partial-wave unitarity requirement $\lambda\le2\pi$ (the largest $s$-wave eigenvalue in the $h_Dh_D$ and longitudinal-$Z'$ (Goldstone) channels is $\lambda/4\pi$, and we require $|{\rm Re}\,a_0|\le1/2$ as above), we obtain
\begin{equation}
C\ge\frac{y^4}{\pi(m_2-m_1)^2}.
\label{a:eq:Cmin}
\end{equation}
We adopt this minimum self-scattering strength to derive the most conservative bound on $y$ within the dark-Higgs completion. For a general quartic coupling, $C=2y^4/[\lambda(m_2-m_1)^2]$, so that $y_c\propto\lambda^{1/4}$ and $g_{\nu,90}\propto1/y_c\propto\lambda^{-1/4}$. The choice $\lambda=2\pi$ is thus the most conservative for $y$ but the most favorable for the limit on $g_\nu$, by a factor of $(2\pi)^{1/4}\simeq1.6$ relative to $\lambda=1$.

The expected number of self-scatterings per DM particle over the black-hole age $t_{\rm BH}$ is~\cite{Shapiro:2014oha,Tulin:2017ara}
\begin{equation}
N(r)=\rho_\chi(r)\frac{\sigma_V}{m_1}
v_{\rm rel}(r)t_{\rm BH},
\end{equation}
where $v_{\rm rel}\simeq\sqrt2\,v_c$ and $v_c=(R_S/2r)^{1/2}$ is the local circular velocity. Requiring the spike to remain collisionless at the neutrino emission radius, $N(R_{\rm em})<1$, gives
\begin{equation}
\begin{aligned}
y&<y_c\equiv
\left[\pi(m_2-m_1)^2C_{\max}\right]^{1/4},\\
C_{\max}&=
\left[
\frac{12\pi}
{m_1\rho_\chi(R_{\rm em})v_{\rm rel}t_{\rm BH}}
\right]^{1/2}.
\end{aligned}
\label{a:eq:yc}
\end{equation}
Consequently,
\begin{equation}
g_\chi<y_c\frac{m_{Z'}}{m_2-m_1}
\simeq y_c\sqrt\epsilon.
\end{equation}
The constraint depends only weakly on the spike properties, scaling as $y_c\propto(\rho_\chi t_{\rm BH})^{-1/8}$. For the NGC~1068 benchmarks, we find $y_c\ll y_{\max}$, demonstrating that spike self-consistency provides the dominant restriction on the dark coupling.

\section{Limits versus kernel width}
\label{app:kernelwidth}

The energy loss in each inelastic scattering exceeds the kinematic minimum $\Eth$ by $Q^2/(2m_1)$, with the characteristic momentum transfer set by $Q^2\sim m_{Z'}^2$. The relative width of the energy-loss distribution is therefore controlled by $\epsilon=m_{Z'}^2/(m_2^2-m_1^2)$. Here we investigate how the constraints on the asymptotic optical depth $\tau_0$ depend on this kernel width.

\begin{figure}[htbp]
\centering
\includegraphics[width=0.95\textwidth]{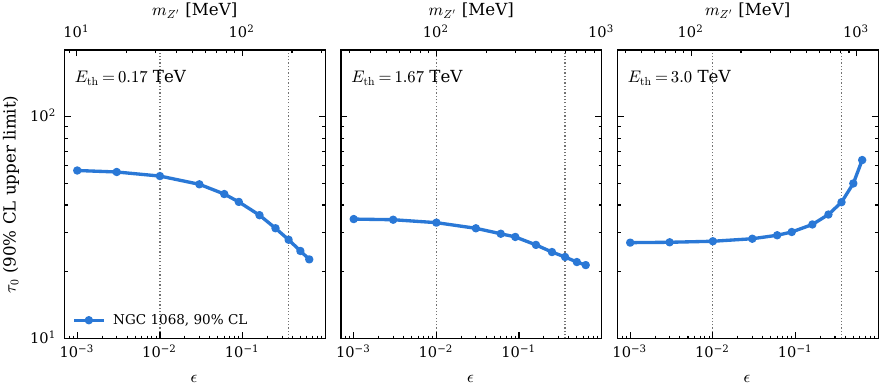}
\caption{Pointwise upper limits on the asymptotic optical depth $\tau_0$ from the NGC~1068 likelihood analysis as functions of the kernel width $\epsilon$, for fixed thresholds $\Eth=0.17$, 1.67, and 3.0~TeV (nominal 90\% CL). The dotted lines indicate the benchmark values $\epsilon=0.01$ and 0.36. The upper axis shows the corresponding mediator mass $m_{Z'}=\sqrt{2\epsilon m_1\Eth}$ for $m_1=300$~keV.}
\label{a:fig:epsscan}
\end{figure}

Figure~\ref{a:fig:epsscan} shows the NGC~1068 likelihood limits (Sec.~\ref{sec:likelihood}) as functions of $\epsilon$ at three fixed thresholds, $\Eth=0.17$, 1.67, and 3.0~TeV, scanning $\epsilon=0.001$--0.64. This range satisfies the kinematic condition $m_{Z'}<m_2-m_1$, ensuring that the on-shell decay $\chi_2\to\chi_1Z'$ remains allowed and that the secondary-neutrino treatment applies. For $\epsilon\lesssim0.01$, the limits approach the narrow-kernel regime, varying by at most 6\% between $\epsilon=0.001$ and 0.01.

At $\Eth=0.17$ and 1.67~TeV, increasing $\epsilon$ strengthens the constraints, with the upper limits on $\tau_0$ decreasing by factors of 2.4 and 1.6, respectively, as $\epsilon$ increases from 0.01 to 0.64. This behavior arises because broader kernels produce larger energy losses per scattering, enhancing the depletion of the neutrino flux. At $\Eth=3.0$~TeV, however, the trend reverses: the upper limit weakens by a factor of 2.3 over the same range. In this regime, the broadened spectral edge overlaps the energies most relevant to the likelihood, allowing greater transmission immediately above threshold and reducing the sensitivity to attenuation. At common threshold values, these results are consistent with the corresponding fixed-width scans shown in Fig.~\ref{fig:constraints} for $\epsilon=0.01$ and 0.36.

\bibliography{refs}
\end{document}